# Observation of flat-band skin effect

Xulong Wang[1,*], Dongyi Wang[1,*], Congwei Lu[1,*],
Ruo-Yang Zhang[2], Ching Hua Lee[3], Kun Ding[4], Guancong Ma[1,5,†]

[1]Department of Physics, Hong Kong Baptist University, Kowloon Tong, Hong Kong, China

[2]Department of Physics, Nanjing University, Nanjing, China

[3]Department of Physics, National Singapore University, Singapore 117542, Singapore

[4]Department of Physics, State Key Laboratory of Surface Physics, and Key Laboratory of Micro and Nano Photonic Structures (Ministry of Education), Fudan University, Shanghai 200438, China

[5]Shenzhen Institute for Research and Continuing Education, Hong Kong Baptist University, Shenzhen 518000, China

## Abstract

Symmetry-protected ideal flat bands in one-dimensional (1D) Hermitian lattices are populated by compact localized states (CLS) – a special class of localization with wavefunctions confined within a small region. In this work, we discover that the non-Hermitian skin effect (NHSE) can appear in a flat band. Unlike conventional NHSEs for dispersive bands that are protected by nontrivial point-gap topology, the flat-band skin effect (FBSE) emerges even though the flat band remains a point on the complex-energy plane with trivial spectral topology. The FBSE is intriguingly associated with the non-trivial spectral topology of the dispersive bands enclosing the flat band, so it only emerges within a finite range of non-Hermitian parameters and can counterintuitively disappear at large non-Hermiticity. We experimentally observe the FBSE and its parameter dependence in a non-Hermitian mechanical lattice. Our theoretical analyses further reveal that the gap between the flat and the dispersive bands can close at higher-order exceptional points (EPs) under both periodic and open boundary conditions. The flat-band wavefunctions are discontinuous in quantum distance across these EPs, a property only found in singular flat bands in Hermitian systems. Our work reveals flat-band phenomena unique to non-Hermitian systems and highlights new possibilities in quantum geometry and localization control.

* These authors contributed equally to this work.

† Email: phgcma@hkbu.edu.hk

*Introduction*—Bragg scattering of waves by a periodic potential generates energy bands that are usually dispersive, with the slope representing the group velocity of a wave. When specific symmetry conditions are met, bands with energy independent of Bloch momentum can appear. These are called flat bands [1-5]. Waves or carriers on a flat band have vanishing group velocity, resulting in extremely low mobility – a condition that benefits interactions which are essential for many-body phenomena such as ferromagnetism [6,7], superconductivity [8-11] and fractional quantum Hall effects [12-16]. The vanishing group velocity is also an ideal condition for the manifestation of quantum-geometric phenomena [10,17,18]. Consequently, the research on flat bands has received tremendous attention in condensed matter physics [8,19-22]. Flat bands are also useful for photonics. For example, the divergent density of states can lead to non-Markovian radiation for quantum emitters [23], and the highly degenerate modes are candidates for lasing devices [24,25]. For a flat band isolated by bandgaps, under a proper choice of basis, the wavefunctions form compact localized states (CLS) with a highly confined support [26,27]. Unlike other types of localization characterized by wavefunctions with decaying tails, CLS are non-zero within a small region and are exactly zero elsewhere. Because the CLS are $\mathcal{O}(L)$-fold degenerate ($L$ being the size of the specimen), they play a fundamental role in many novel physical phenomena, such as non-Abelian dynamics [28-32].

On a different frontier, non-Hermitian physics, a theoretical formalism that judiciously exploits energy or particle exchange with external reservoirs, brings revolutions across many realms of physics and engineering [33-42]. In particular, the crossover of non-Hermitian formalism with topological matter results in the discovery of the non-Hermitian skin effect (NHSE), by which eigenmodes of dispersive bands in non-Hermitian periodic systems become skin modes localized at open boundaries [43-46]. Intriguingly, the interplay between NHSE and other localization mechanisms results in rich transport phenomena. Examples include the delocalization of bound states protected by line-gap topology [47-49], the suppression of Landau quantization induced by effective magnetic flux [50,51], as well as Anderson transitions and NHSE-induced criticality [52] in one-dimensional (1D) disorder or impurity lattices [53-55]. Although previous works have investigated non-Hermitian 1D flat-band models with a focus on the topological properties [56-58] and their engineering through gain-loss modulation and specific lattice symmetries [59-61], the interplay between NHSE and flat bands populated by CLS remains largely unexplored. In this work, we study 1D non-Hermitian lattices that sustain an ideal flat band. We find that the flat-band modes are affected by NHSE under an open boundary condition (OBC). But unlike NHSEs for usual dispersive bands that originate from the nontrivial spectral topology of the same bands, the flat-band skin effect (FBSE) exists only when the point gap of the dispersive bands encloses the flat band on the spectral plane under the periodic boundary condition (PBC). As such, FBSE can counterintuitively disappear under large non-Hermitian parameters. More intriguingly, the non-Hermitian parameters can close the OBC gaps between the flat band and dispersive bands in the formation of order-3 exceptional points (EP3s), and the gap-closing point is singular on the non-Bloch wavevector plane. These phenomena go beyond the common understanding that flat bands in 1D systems are non-singular [3,5]. We used an active mechanical lattice to experimentally demonstrate the FBSE. Our findings reveal intriguing non-Hermitian phenomena associated with flat bands and may break new ground for metamaterials, photonics, and condensed matter physics.

*Non-Hermitian flat band model*—Consider an $(M+1)$-dim square matrix in the form of $H=\begin{pmatrix} 0 & A_{1\times M} \\ B_{M\times 1} & 0_{M\times M}\end{pmatrix}$, where $A_{1\times M}$ and $B_{M\times 1}$ are non-zero vectors, and the subscripts indicate dimensions. Regardless of $M$, the matrix is always $\mathrm{rank}(H)=2$. In other words, $H$ has an $(M-1)$-dim null space "spanned" by degenerate eigenvectors with a zero eigenvalue. These zero modes exist even when $H\neq H^{\dagger}$. This construction route was discussed as a general class of bipartite crystalline lattices recently [61]. Construct a three-band non-Hermitian Bloch Hamiltonian with such a structure

$$H(k)=\begin{pmatrix} 0 & t_1+\gamma_1+t_2e^{-\mathrm{i}k} & t_2+\gamma_2 \\ t_1-\gamma_1+t_2e^{\mathrm{i}k} & 0 & 0 \\ t_2-\gamma_2 & 0 & 0 \end{pmatrix}, \tag{1}$$

where $k$ is the Bloch wavevector, and $t_1,t_2,\gamma_1,\gamma_2$ are real-valued parameters representing reciprocal and non-reciprocal hoppings, respectively. A schematic drawing of the lattice is shown in Fig. 1(a). Apparently, $M=2$ for Eq. (1), so there is one band with zero eigenenergy for all $k$. This is the system's flat band. Meanwhile, owing to the presence of non-reciprocal hoppings, the two dispersive bands form closed loop(s) on the complex energy plane, indicating the presence of NHSE when the lattice is under an OBC. Importantly, the existence of the flat band is due to the mathematical structure of the Hamiltonian and is unaffected by the non-Hermiticity [62]. It is always a single point at the origin of the complex energy plane, as shown in Fig. 1(c1-4).

A notable characteristic of Model (1) is that the gaps between the flat band and dispersive bands do not close under the Hermitian case. This can be easily verified by setting $\gamma_1 = \gamma_2 = 0$ in Eq. (1): the dispersive bands are given by $E(k) = \pm\sqrt{\Delta}$ with $\Delta = t_1^2 + 2t_2^2 + 2t_1 t_2 \cos k$, which vanish only when $t_1 = t_2 = 0$, at which point the Hamiltonian itself vanishes, i.e., $H(k) = 0$. However, this is not the case when the model becomes non-Hermitian with $\gamma_1, \gamma_2 \neq 0$. The dispersive bands are $E(k) = \pm\sqrt{\Delta - (\gamma_1^2 + \gamma_2^2) + 2\mathrm{i}\gamma_1 t_2 \sin k}$, and the flat band remains $E(k) = 0$. So the line gaps close at $k = \pm\pi$ when $t_1^2 + 2t_2^2 - 2t_1 t_2 = \gamma_1^2 + \gamma_2^2$ and re-open at $k = 0$ when $t_1^2 + 2t_2^2 + 2t_1 t_2 = \gamma_1^2 + \gamma_2^2$. These two distinct conditions correspond to the boundaries of the non-Hermitian phase transition. The PBC gap-closing positions are indicated in Fig. 1(b) as the red curves. This gap-closing point is an EP3. (In a standard non-Hermitian SSH model, the PBC gap-closing point is a second-order EP.) The PBC spectra under different $\gamma_2$ are plotted in Figs. 1(c1-4).

*Revealing the FBSE*—Let us examine the OBC eigenmodes on the flat band. These flat-band modes are $N$-fold degenerate, so they can form numerous wavefunction profiles through linear combinations. To properly determine the spatial characteristics of the flat-band modes, it is necessary to examine a basis-invariant quantity. The Green's function of the system, $\hat{G}(E_\eta) = (E_\eta - H_{\mathrm{OBC}})^{-1}$, is a good choice. Here,

$$\begin{aligned} H_{\mathrm{OBC}} = \textstyle\sum_{n=1}^{N} \big[ &(t_1+\gamma_1) a_n^\dagger b_n + (t_1-\gamma_1) b_n^\dagger a_n + (t_2+\gamma_2) a_n^\dagger c_n + (t_2-\gamma_2) c_n^\dagger a_n \big] \\ &+ \textstyle\sum_{n=1}^{N-1} t_2 \big( b_n^\dagger a_{n+1} + \mathrm{h.c.} \big) \end{aligned} \tag{2}$$

is the OBC Hamiltonian, which has the same algebraic structure as Eq. (1) after a unitary transformation,and $E_\eta = 0 + \mathrm{i}\eta$ with $\eta \to 0$ is an energy close to zero ($\eta = 10^{-20}$ was used in our numerical calculations.), so that the source selectively excites the flat-band modes. We set $H_{\mathrm{OBC}}$ to be gapped such that any source with an energy $E_\eta$ only excites the flat-band modes. We examine the normalized spatial responses $|R\rangle = \hat{G}(E_\eta)|s\rangle$, where $|s\rangle$, representing the source profile, is a $3N \times 1$ normalized column vector with only one non-zero entry at site C of the first unit cell [indicated in Fig. 1(a)]. We further cast the normalized $|R\rangle$ into a quantity carrying a similar meaning as the center of mass of a rigid body:

$$\chi \coloneqq \sum_{n=1}^{3N} \frac{n}{3N} |R_n|^2 \tag{3}$$

where $R_n$ is the entry of $|R\rangle$ at site-$n$. If the flat-band modes remain localized across the bulk, $\chi$ is almost zero because the excitation position is at the leftmost unit cell. If the FBSE is present, $\chi$ approaches unity. In Fig. 1(b), we plot $\chi$ as a function of the two non-Hermitian parameters $\gamma_1$ and $\gamma_2$. It can be seen that $\chi$ is near unity in a well-defined region, signifying that the response is localized at the right edge. Clearly, the FBSE exists in this region. The existence and absence of the FBSE are further confirmed by the exponential scaling of $\max G(E_\eta)$ with respect to lattice size, as shown in the Supplemental Material [63]. Further analysis confirms the robustness of this boundary-localized response against the choice of source profile, excitation position, and the $E_\eta \to 0$ limit is discussed therein.

To better understand these results, we examine the OBC flat-band modes. These modes are degenerate at zero energy, they span an $N$-dim null space, denoted $\ker H_{\text{OBC}}$. Despite being non-Hermitian in character, the null space can still be decomposed by a CLS basis [see Eqs. (5) and (6) in the End Matter], such that the right eigenvectors (REVs) are CLS across the bulk even in the FBSE regime, as shown in Fig. 2. The choice of this basis is not unique, and the compact REVs forming the CLS basis are non-orthogonal to one another [64]. For instance, directly performing standard orthogonalization procedures, such as a QR decomposition on the null space, yields a set of slightly distinct REVs. From this perspective, it is surprising that the FBSE dominates the system's responses [region II in Fig. 1(b)]. To understand this, we write the flat-band Green's function in the Lehmann spectral representation, i.e., $G(E_\eta) = \sum_n E_\eta^{-1} |\psi_{0,n}^R\rangle\langle\psi_{0,n}^L|$, where both REVs and their biorthogonal dual, left eigenvectors (LEVs), play an equal role [65]. We compute the LEVs from the set of REVs. In regions I and III where FBSE is absent, all LEVs are localized at the same position as the corresponding REVs [Figs. 2(a) and (c)]. However, in region II, a small number of LEVs are exponentially localized at the left boundary with excessively large entries [Fig. 2(b)]. In contrast, the corresponding REVs are localized at the opposite boundary [66]. Clearly, these LEVs are the reason why FBSE dominates the system's response and manifesting as a biorthogonal-response effect.

The FBSE exhibits several unexpected characteristics. First, it deviates from the standard non-Hermitian bulk-boundary correspondence [67-69]: under PBC, the flat band always remains as a single point at the origin – its PBC spectrum has trivial point-gap topology and has no interior in which the skin modes can lie. Second, unlike NHSE for dispersive bands, the FBSE only exists for a well-defined range of $\gamma_1, \gamma_2$ [region II in Fig. 1(b)], and it counterintuitively disappears under large values (region III). The disappearance of the FBSE under large non-Hermiticity is a disparate feature not seen for NHSEs in dispersive bands. By examining the PBC and OBC spectra of the system under different parameters [marked in Fig. 1(b)], as shown in Fig. 1(c1-4), we find that the FBSE region coincides with the region where the line gaps between the dispersive bands are closed, which is marked by the red curves in Fig. 1(b). As such, the FBSE is surprisingly linked to the point-gap topology of the two dispersive bands: the FBSE emerges when the flat band falls inside the point gap formed by the dispersive bands under PBC Under large non-Hermiticity, the line gap reopens between the dispersive bands, so the flat band is again outside any spectral loop, and the FBSE disappears. These observations suggest that the FBSE originates from, rather surprisingly, the nontrivial point-gap topology of the surrounding dispersive bands. An analytical derivation linking the FBSE with the winding of the dispersive bands via Gram matrices is presented in the End Matter.

In Fig. 1(b), there is a horizontal line $\chi \cong 0$, indicating the absence of FBSE. This line is at $\gamma_2 = -t_2 = 0.3$, i.e., the hopping between site C and the main chain becomes purely unidirectional, so the response $|R\rangle$ only localized at source (site C in the first unit cell).

*Experimental observation of the FBSE*—We experimentally realize the OBC model using an active mechanical lattice [Fig. 3(a)], an in-house developed platform for faithful and versatile realizations of Hermitian and non-Hermitian systems [47,48,54,70-73]. Simply put, the lattice has 36 spring-coupled rotational harmonic oscillators (12 unit cells), and the non-reciprocal coupling is enabled by programmable brushless DC motors driven by active feedback. (See the Supplemental Material for experimental details). All oscillators are carefully tuned to an identical resonant frequency $f_0 = 10.9$ Hz, which is also the experimental frequency of the flat band.

We demonstrate the FBSE by probing the steady-state response to a local harmonic excitation at site C with frequency $f_0$. We first consider parameters in region I [at $(\gamma_1, \gamma_2) = (0.62, 0.32)$, marked by the grey arrow in Fig. 1(b)], where the FBSE is absent. The responses are localized around the source [Figs. 3(b1, c1)]. When the parameters are shifted to region II [$(\gamma_1, \gamma_2) = (0.9, 0.32)$, blue arrow in Fig. 1(b)], the responses are no longer localized around the source but at the right edge, regardless of the excitation positions [Figs. 3(b2, c2)]. These results confirm the FBSE. We further compare the FBSE with the NHSE of the dispersive bands. The responses in the dispersive band show clear NHSE across all parameters [Figs. 3(d) and (e)].

A unique feature of the FBSE is its vanishment when the flat band lies outside the point gap of the dispersive bands under large non-Hermitian parameters. In this region, the OBC dispersive bands have undergone a non-Bloch PT transition, so all modes have a significant imaginary part in their eigenenergies [Fig. 1(c4)]. One dispersive band exclusively contains gain modes, and some of these modes also have near-zero real energies. The response of these modes, which are spatially edge-localized and temporally divergent, will dominate over that of the flat band, preventing the proper observation of the vanishing of the FBSE. To overcome this issue, we implement a transformed model that is eigenstate-equivalent to the model in Eq. (2) but with a spectrum rotated by $90°$ on the complex plane. The mathematics of the transformations is presented in the End Matter. The dispersive bands and the flat band in the transformed lattice now have different real energies, so the flat-band modes can be excited without the influence of the gain modes in the dispersive band by using a mono-frequency source. The measured results [in region III at $(\gamma_1, \gamma_2) = (1.5, 0.66)$] show that the responses are localized near the source [Fig. 3(b3)], confirming the absence of the skin effect.

*EP3s on the generalized Brillouin zone (GBZ)*—We further analyze the non-Hermitian flat band using the GBZ, a theoretical tool that reconciles the PBC and OBC eigenstates by complex-deforming the Bloch wavevector [43,74], i.e., replacing $e^{ik}$ with $\beta \in \mathbb{C}$, where $\beta$ is the non-Bloch wavevector. To obtain the GBZ, the general practice is to solve $\det[H(\beta) - E_n] = 0$, where $E_n$ represents the energy of the $n$-th OBC eigenstate. The GBZs for the two dispersive bands are two identical circles with $|\beta| > 1$. However, because $E = 0$ for the flat band, the equation reduces to $\det H(\beta) = 0$, which always holds for our model. In general, such flat bands from rank-defective Hamiltonians always admit an unconventional GBZ that encompasses all values on the complex-$\beta$ plane.

In Figs. 4(a1-a4), we plot the GBZs together with the eigenenergies of $H(\beta)$ at different parameters. It is seen that the GBZs of the two dispersive bands can merge with the flat band at zero energy. The merging points are a pair of EP3s, as indicated by the red dots in Fig. 4(a2, a3). Further increasing $\gamma_2$ drives the two EP3s to move along the GBZs of the dispersive bands, resembling two “zippers” that close (open) for the real (imaginary) eigenenergies along the GBZs (see the Supplemental video). Throughout this evolution, the GBZs and the positions of the EP3s are found at $|\beta| > 1$. Eventually, the EP3s disappear at large non-Hermiticity. More discussion on EP3s and their observable consequences in Supplemental Material [63].

The EP3s on the GBZ possess important characteristics of the singular gap-closing point. In the Hermitian context, the Bloch eigenstates at the gap-closing point of a singular flat band are discontinuous, and the CLS do not form a complete set of basis [3,5]. Here, because the gap-closing points are EP3s, i.e., branch singularities on the spectrum, it is not possible to continuously evolve an eigenstate across them. We compute the quantum distance on the complex-$\beta$ plane in the vicinity of the EP3. In a similar spirit to non-Hermitian topological invariants [75], there are two possibilities in defining the quantum distance in a non-Hermitian context:

$$d_{LR}(\theta, \Delta\theta)|_{\delta\beta} = \sqrt{\left|1 - \left|\langle \bar{\psi}_0^L(\theta + \Delta\theta) | \bar{\psi}_0^R(\theta) \rangle\right|^2\right|},$$

$$d_{RR}(\theta, \Delta\theta)|_{\delta\beta} = \sqrt{\left|1 - \left|\langle \bar{\psi}_0^R(\theta + \Delta\theta) | \bar{\psi}_0^R(\theta) \rangle\right|^2\right|} \tag{4}$$

where $\beta = |\beta| e^{\mathrm{i}\theta}$, $\Delta\theta = \arg(\beta - \beta_{EP})$, $\delta\beta \coloneqq |\beta - \beta_{EP}|$, $\langle \bar{\psi}_0^L|$ and $|\bar{\psi}_0^R\rangle$ ($\langle \bar{\psi}_0^R| = |\bar{\psi}_0^R\rangle^\dagger$) are the biorthogonal normalized LEV and REV of the non-Bloch zero mode. Figures 5(a) and (c) plot the $d_{LR}$ and $d_{RR}$ at $\delta\beta = 10^{-3}$, respectively. It is seen that $d_{RR}$ is isotropic in all directions but $d_{LR}$ varies with $\Delta\theta$. In addition, $d_{RR}$ vanishes when $\delta\beta \to 0$, but $d_{LR}$ remains finite [Figs. 5(b) and (d)]. The behaviors of $d_{LR}$ carry essential characteristics of the singular gap-closing point [5,18], which are not expected for 1D Hermitian flat bands. A directionally dependent (anisotropic) $d_{LR}$ directly metrics the singular discontinuity of dual space, associated with EP3. It serves as the definitive geometric signature that this is a branch singularity rather than a trivial degeneracy.

Because the GBZs are found by matching the OBC spectrum with the non-Bloch Hamiltonian $H(\beta)$, the EP3s also appear in the OBC spectrum. The EP3s form continuous curves on the $\gamma_1\gamma_2$-plane, as shown in Fig. 4(b1, b2), and their number scales with the OBC lattice size. The EP3s are also a consequence of the FBSE—the three eigenmodes are parallel, which is only possible when the flat-band modes can produce a wavefunction identical to the two coalescing skin modes from the dispersive bands. Intriguingly, although the flat-band modes are $N$-fold degenerate, they only supply one eigenmode to coalesce with a pair of modes from the two dispersive bands, whereas the remaining $N-1$ modes on the flat band remain linearly independent. At such points, the algebraic multiplicity of the OBC Hamiltonian is $N+2$, but the geometric multiplicity is $N$ [76-80]. The verification was performed symbolically for small $N$ or analytically by deriving the Jordan blocks of the Hamiltonian's canonical form, as detailed in the End Matter. Detailed mathematics accounting for the order of the EPs and the emergence of flat-band skin modes is presented in the End Matter. Clearly, at the EP3, the flat-band mode must possess the same $\beta$ as two dispersive bands, so that they can coalesce.

*Conclusion*—Our work unveils a new type of NHSE appearing uniquely in ideal flat bands. As directly demonstrated in our experiments, unlike standard NHSE in dispersive bands, the FBSE exhibits a distinct re-entrant behavior: it emerges upon PBC gap-closing and counterintuitively vanishes under large non-Hermitian parameters when the line gaps re-open. The FBSE, originating from the nontrivial spectral topology of the dispersive bands, is an unconventional manifestation of non-Hermitian spectral topology and has been realized in the mechanical lattice. Furthermore, our theoretical GBZ analysis reveals that the non-Bloch bandgaps between the flat and dispersive bands can close through the formation of EP3s, which leads to discontinuous non-Hermitian quantum distances. Future investigations can further explore the possible effects hinged on the non-Hermitian quantum geometry [8,81] and its experimental validation or extend to two-dimensional non-Hermitian flat-band models, in which the possible emergence of loop states of a non-Hermitian origin is a particularly interesting question. Our results are also potentially useful for new photonic applications.

*Acknowledgement*—This work was supported by supported by the National Natural Science Foundation of China (NSFC, T2525002), the National Key R&D Program (2022YFA1404400), the Hong Kong Research Grants Council (RFS2223-2S01, 12301822, 12300925), and the Hong Kong Baptist University (RC-RSRG/23-24/SCI/01, RC-SFCRG/23-24/R2/SCI/12). K. D. acknowledges the support from the NSFC (12174072, 2021hwyq05). C. H. L. acknowledges support from the Ministry of Education, Singapore (MOE award number: MOE-T2EP50222-0003). X. W. was also supported by the Hong Kong RGC Junior Research Fellowship (JRFS2526-2S07).

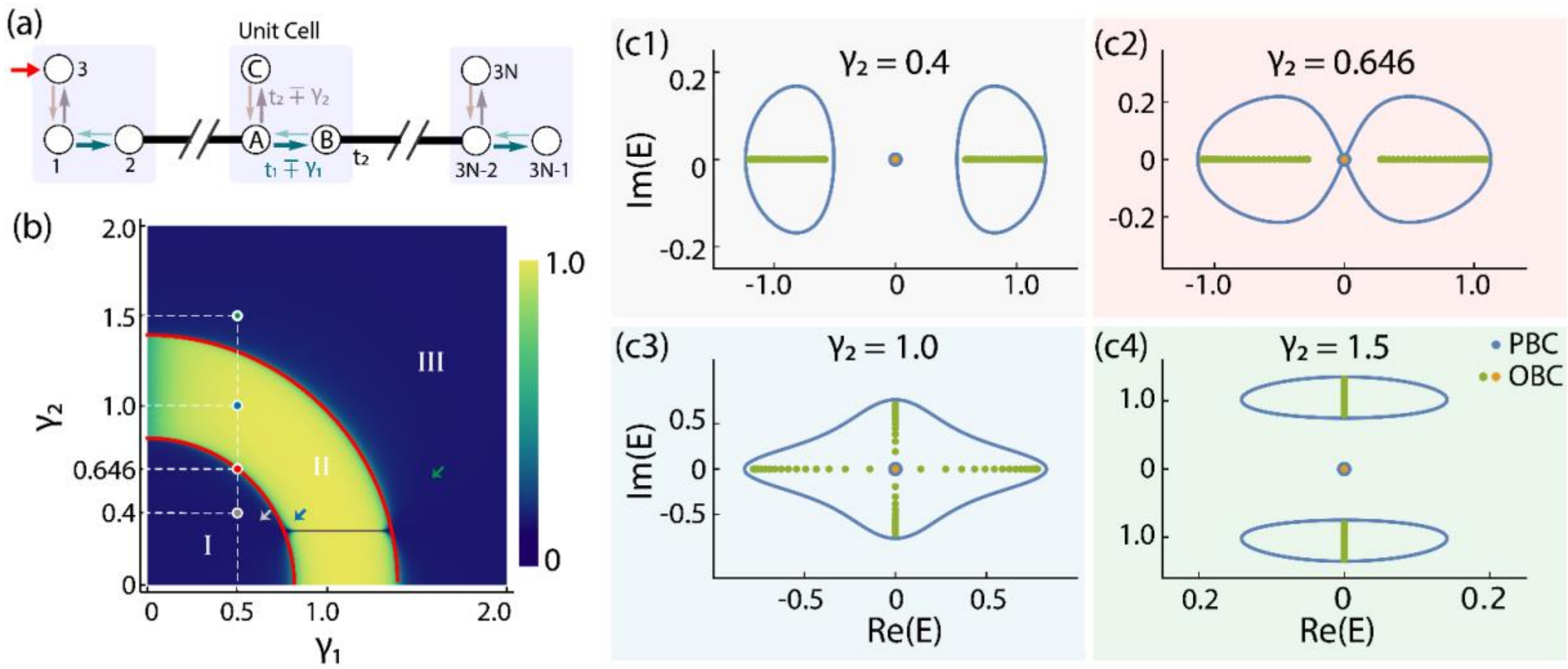

FIG. 1. The emergence of the flat band skin effect (FBSE). (a) The 1D non-Hermitian flat-band model [Eq. (1)]. (b) The averaged responses $\chi$ of the OBC flat band computed through the Green's function. The red curves mark the closing and opening positions of PBC line gaps. The FBSE appears in the region between the red curves, i.e., when the flat band is inside the point gap. The colored dots mark representative parametric positions chosen for the investigations in (c). The arrows (gray, blue and green) indicate the experimental parameter combinations used in our experimental setups 1 to 3. (c1-c4) PBC and OBC spectra of the flat band model under various $\gamma_2$ and fixed $\gamma_1 = 0.5$, as indicated by the correspondingly colored dots in (b). The OBC lattice has 20 unit cells. Other parameters are $t_1 = -1.06$ and $t_2 = -0.3$.

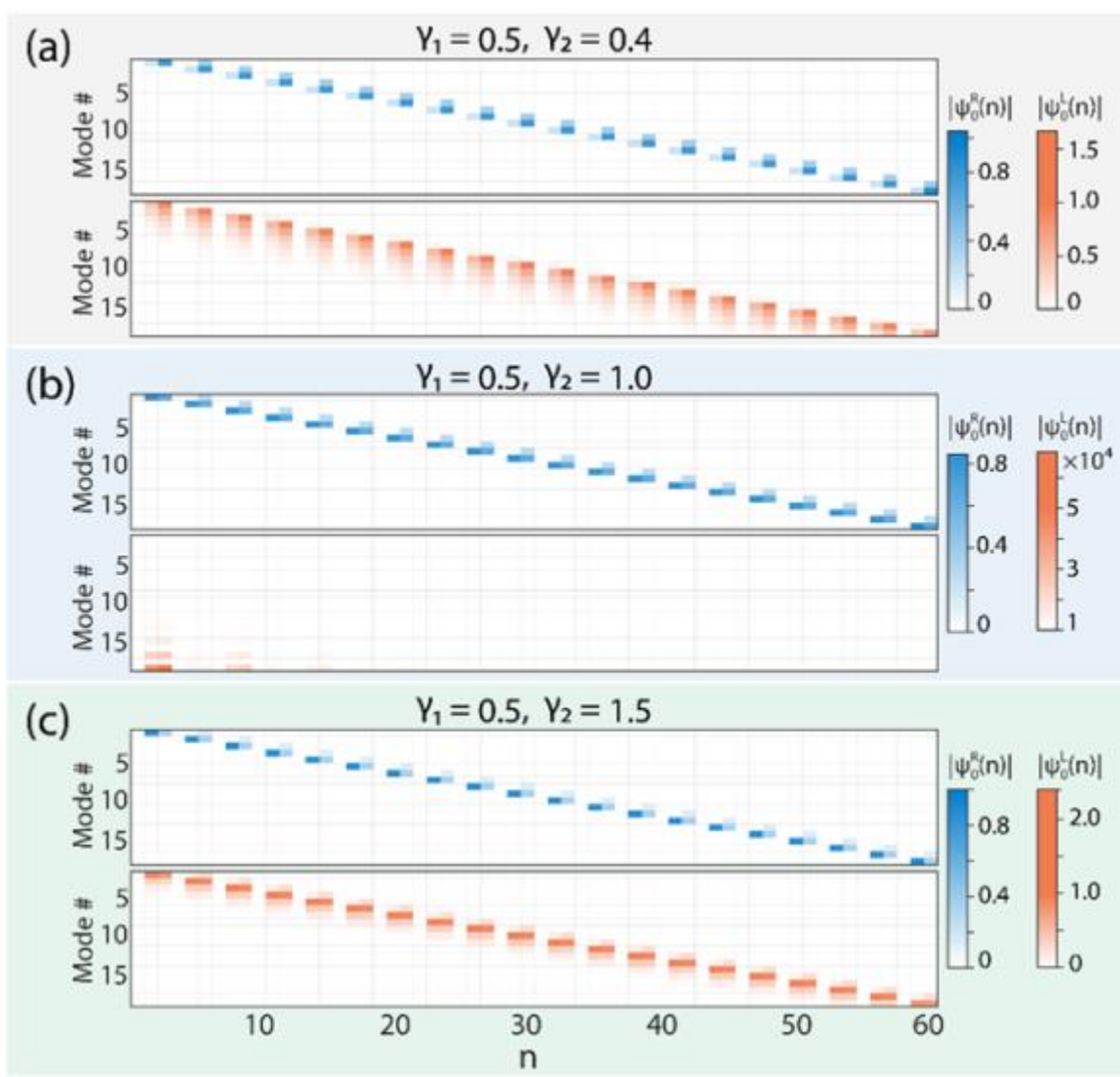

FIG. 2. Decomposition of the REVs of OBC flat-band modes by a CLS basis. (a, c) When the PBC flat band lies outside the point gaps, the REVs are CLS and the corresponding LEVs are also localized across the lattice. Non-Hermitian skin modes are absent. (b) While the REVs remain as CLS uniformly distributed across the lattice even in the FBSE regime, all corresponding LEVs become localized at the left boundary. Notably, the magnitude of these LEVs increases drastically as the center of their paired REVs shifts to the right; the REV at the right edge ($n = 60$) corresponds to a left-localized LEV with divergent amplitudes. This extreme asymmetry signifies the presence of the FBSE.

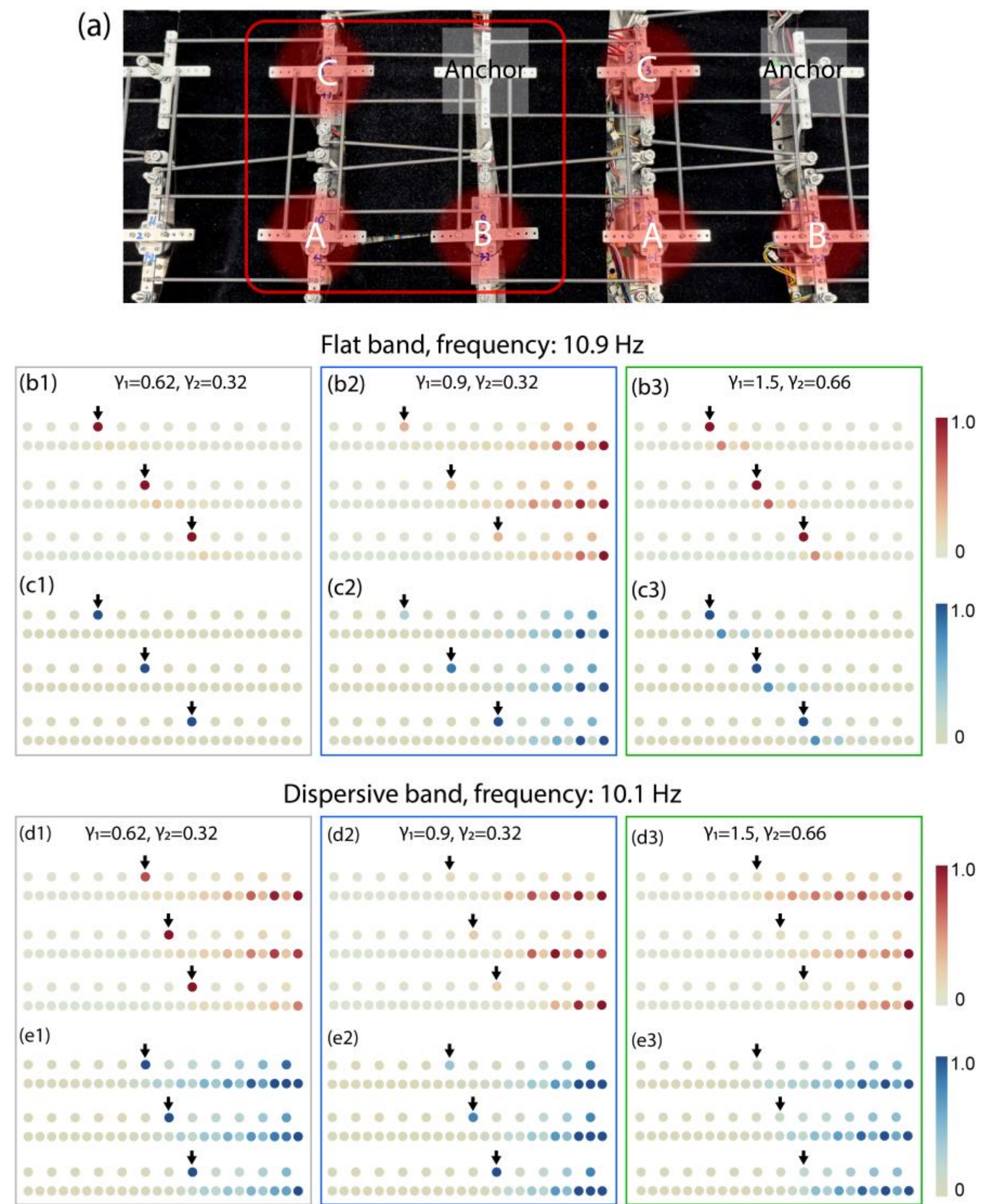

FIG. 3. Experimental observation of the FBSE. (a) The active mechanical lattice realizing the non-Hermitian flat-band model. The red box marks a unit cell. (b, c) Steady-state responses of the flat band ($f = 10.9$ Hz). (b1-b3) The measured normalized angular displacements $|\theta(n)/\theta_{\max}|$. (c1-c3) Steady-state responses $|R(n)|$ computed using the Green's function, showing excellent agreement. The FBSE is observed in the region II (b2, c2), but is absent in regions I and III (b1, b3, c1, c3). (d, e) The responses under an excitation with frequency $f = 10.1$ Hz, which is in the dispersive band. NHSE is observed for all parameters. The arrows indicate excitation sites. In all experiments, $t_1 = -1.06, t_2 = -0.3$ and all effective parameters were extracted by fitting experimental responses to the Green's function. The three parameter sets correspond to the gray, blue, and green arrows in Fig. 1b, respectively.

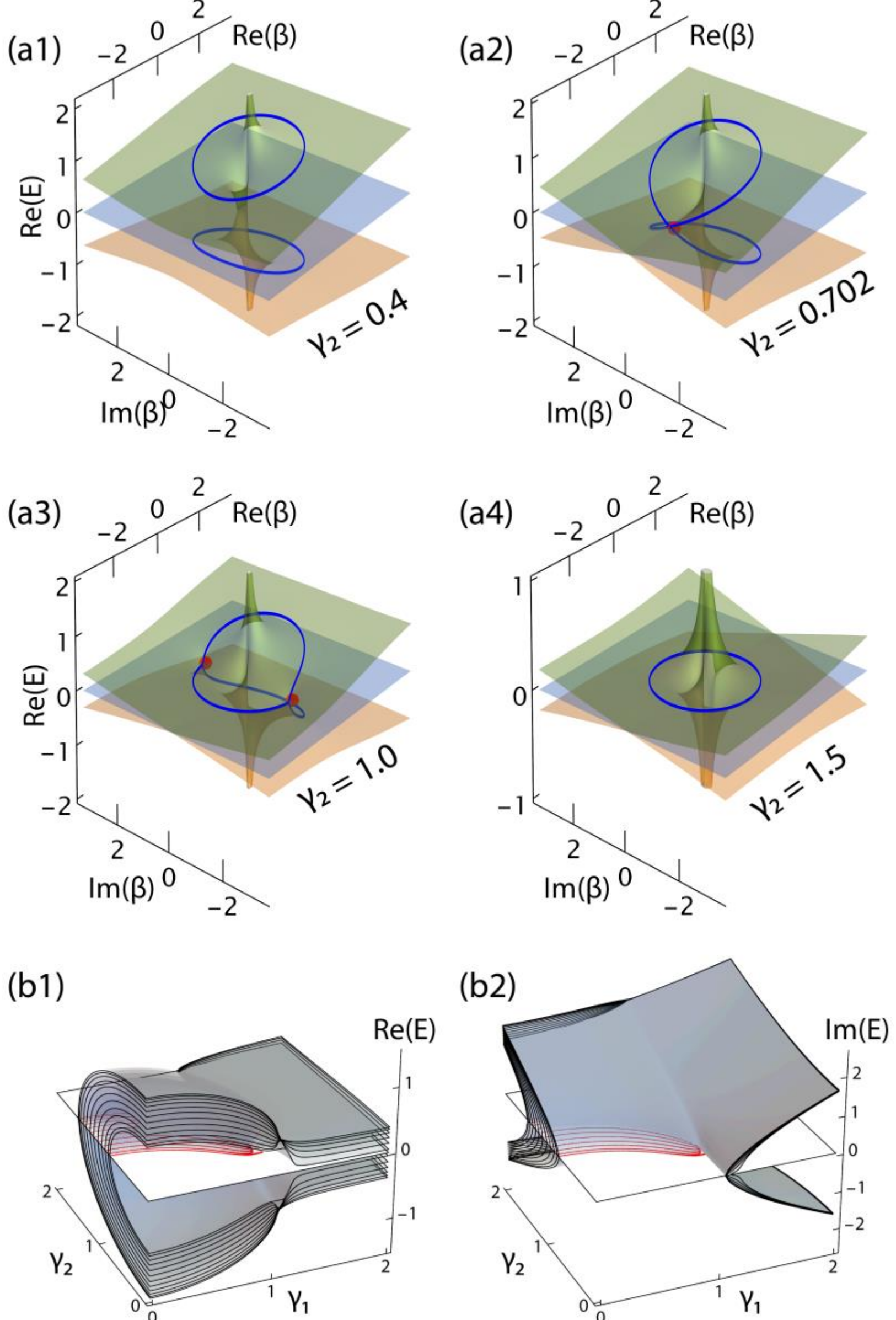

FIG. 4. Non-Bloch GBZ and OBC EP3s. (a1-a4) The GBZs (blue curves) and the EP3s [red dots in (a2, a3)] on the $\beta$-plane. $\gamma_1 = 0.5$ in all panels. The EP3s are formed by the coalescence of one mode from the flat band and two modes from the dispersive bands. (b) The real (b1) and imaginary (b2) parts of the OBC spectrum as functions of $\gamma_1$ and $\gamma_2$. The red curves on the zero-energy plane delineate the EP3s. The OBC lattice has 12 unit cells. Other parameters are $t_1 = -1.06$, $t_2 = -0.3$.

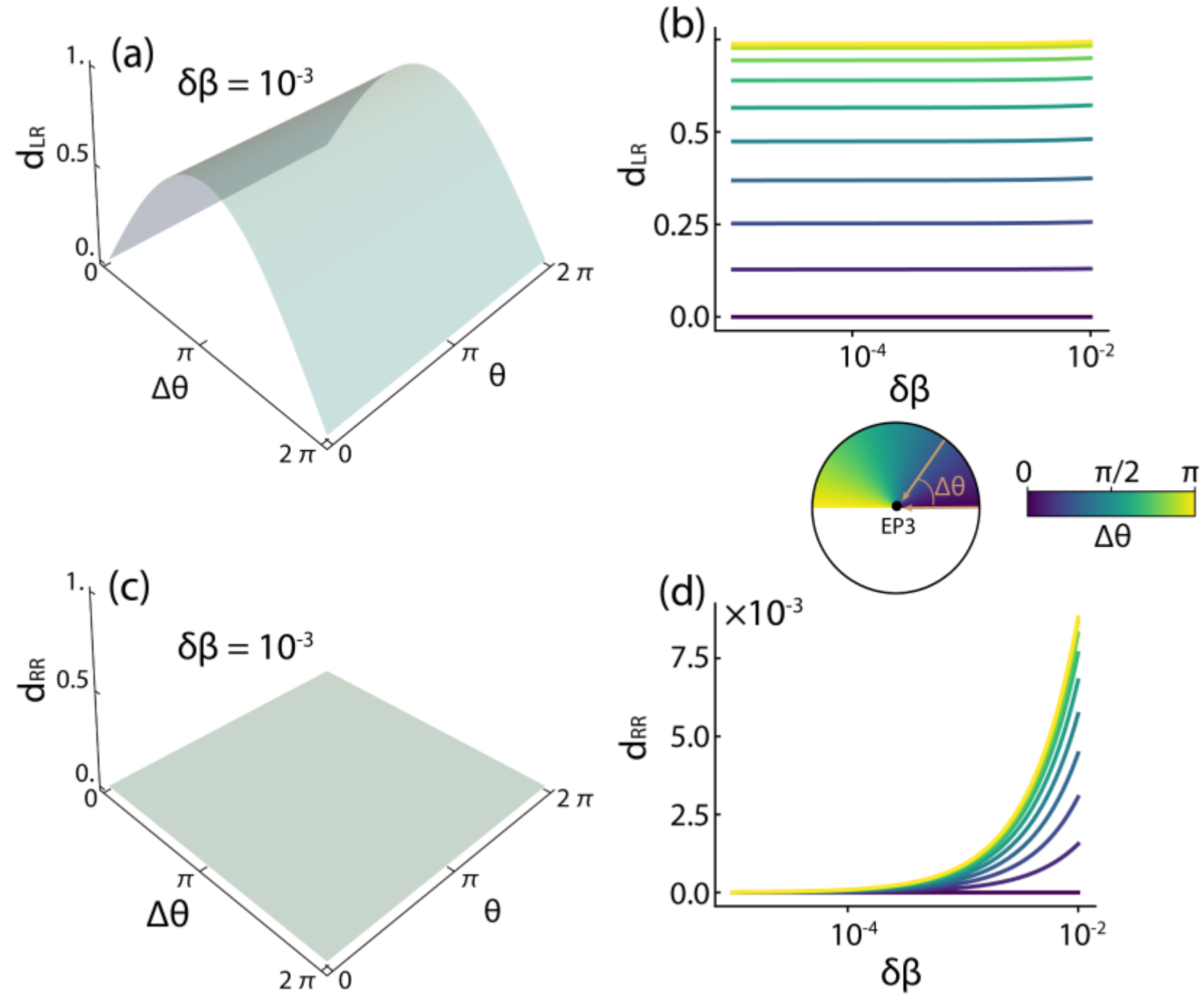

FIG. 5. Non-Hermitian quantum distances near the non-Bloch EP3. (a) The distribution of quantum distances $d_{LR}(\theta, \Delta\theta)$ at $\delta\beta = 10^{-3}$. (b) $d_{LR}$ is different when the EP3 is approached from different directions $\Delta\theta$ (sampled at multiple values, corresponding to different colors) on the $\beta$-plane. (c, d) In comparison, the distribution of quantum distances $d_{RR}$ is isotropic near the EP3. In (b, d), the reference angle $\theta = 0$.

## End Matter

*Winding of the dispersive bands around the flat band and FBSE*—Here, we present the correspondence between the FBSE and the encirclement of the flat band by the dispersive bands. The Bloch REV and LEV of the flat band are $|\psi_0^R(k)\rangle = \left(0, -t_2-\gamma_2, t_1+\gamma_1+t_2e^{-ik}\right)^{\mathrm{T}}$, $\langle\psi_0^L| = \left(0, -t_2+\gamma_2, t_1-\gamma_1+t_2e^{ik}\right)$, respectively. The OBC REVs and LEVs of the flat-band modes can also be obtained using CLS as the basis

$$\begin{cases}\langle\psi_{0,n}^L| = -(t_2-\gamma_2)\langle n,b| + (t_1-\gamma_1)\langle n,c| + t_2\langle n+1,c| \\ |\psi_{0,n}^R\rangle = -(t_2+\gamma_2)|n,b\rangle + (t_1+\gamma_1)|n,c\rangle + t_2|n+1,c\rangle\end{cases} \text{for } n \leq N-1 \tag{5}$$

$$\begin{cases}\langle\psi_{0,n}^L| = -(t_2-\gamma_2)\langle n,b| + (t_1-\gamma_1)\langle n,c| \\ |\psi_{0,n}^R\rangle = -(t_2+\gamma_2)|n,b\rangle + (t_1+\gamma_1)|n,c\rangle\end{cases} \text{for } n = N \tag{6}$$

Crucially, these CLS-form REVs and LEVs do not satisfy biorthogonal normalization, i.e., $\left\langle\psi_{0,m}^L\middle|\psi_{0,n}^R\right\rangle \neq \delta_{mn}$. To obtain the properly biorthogonal normalized LEVs $\left\langle\bar{\psi}_{0,n}^L\right|$, we construct a Gram matrix $\mathcal{G}$ with entries $\mathcal{G}_{mn} = \left\langle\psi_{0,m}^L\middle|\psi_{0,n}^R\right\rangle$. This yields a Toeplitz-like matrix (where only the last diagonal term is perturbed due to the truncation of the lattice)

$$\mathcal{G} = \begin{pmatrix} g_0 & g_l & 0 & 0 \\ g_r & g_0 & g_l & 0 \\ 0 & g_r & \ddots & g_l \\ 0 & 0 & g_r & g_0 - t_2^2 \end{pmatrix}, \tag{7}$$

where $g_0 = t_1^2 + 2t_2^2 - \gamma_1^2 - \gamma_2^2$, $g_l = t_2(t_1+\gamma_1)$ and $g_r = t_2(t_1-\gamma_1)$, so $|g_r| > |g_l|$. Note that the form of $\mathcal{G}$ commonly appears as the OBC Hamiltonian of Hatano-Nelson models. The normalized OBC LEVs are then $\left\langle\bar{\psi}_{0,n}^L\right| = \sum_{m=1}^N (\mathcal{G}^{-1})_{n,m}\left\langle\psi_{0,m}^L\right|$. Since the unnormalized LEVs $\left\langle\psi_{0,m}^L\right|$ are CLS, the profiles of $\left\langle\bar{\psi}_{0,n}^L\right|$ are determined by the corresponding rows of $\mathcal{G}^{-1}$. Note that $\mathcal{G}^{-1}$ is exactly the Green's function of $\mathcal{G}$ under the excitation of a source at zero energy. So, the profiles of $\left\langle\bar{\psi}_{0,n}^L\right|$ are directly determined by whether "NHSE" appears in $\mathcal{G}$ at zero-energy. It is well-established that the NHSE in Hatano-Nelson models originates from the point-gap topology of the corresponding PBC Hamiltonian, which is exactly the Gram function computed by the Bloch REV and LEV here

$$\mathcal{g}(k) = \langle\psi_0^L(k)|\psi_0^R(k)\rangle = g_0 + g_r e^{-ik} + g_l e^{ik} \tag{8}$$

It can be shown that $\mathcal{g}(k) = -E_-(k)E_+(k)$, where $E_\pm(k)$ are the energy of the dispersive bands (see detailed derivation in the Supplemental Material) [63].

In summary, the non-trivial point-gap topology of the dispersive bands with respect to the flat band at zero energy directly determines to the winding of $g(k)$, which dictates the profile of normalized OBC LEVs through $\mathcal{G}^{-1}$. So the FBSE hinges on the flat band being enclosed by the dispersive bands.

*Unitary transformation of the Hamiltonian*—After the reopening of the PBC line gap, the corresponding OBC spectrum of the two continuous bands cease to be purely real—instead they become conjugate pairs each with an imaginary part in the eigenenergy. As such, the system's responses at zero energy are dominated by states with near-zero real energy and large imaginary energy in the upper continuous band, which makes direct observation of the flat band difficult. To circumvent this issue, we perform a transformation to rotate the energy spectrum by 90° on the complex plane, resulting in a new Hamiltonian with a spectrum featuring a line gap symmetric about the imaginary axis. We multiply the original Hamiltonian $H_{\mathrm{OBC}}$ by the imaginary unit $\mathrm{i}$, i.e., $H_1 = \mathrm{i}H_{\mathrm{OBC}}$. Obviously, the eigenvalues of $H_1$ are $\mathrm{i}E$, where $E$ is an eigenvalue of $H_{\mathrm{OBC}}$. However, $H_1$ has purely imaginary entries, which are also difficult to realize experimentally. We further apply a unitary transformation [82] to obtain a purely real Hamiltonian $H_2 = U_1^{\dagger} H'_{\mathrm{OBC}} U_1$, where $H_2 = U_1^{\dagger}[H_1 \oplus (-H_1)]U_1 = \begin{pmatrix} \mathrm{Re}(H_1) & \mathrm{Im}(H_1) \\ -\mathrm{Im}(H_1) & \mathrm{Re}(H_1) \end{pmatrix} = \begin{pmatrix} 0 & H_{\mathrm{OBC}} \\ -H_{\mathrm{OBC}} & 0 \end{pmatrix}$, where $U_1 = \frac{1}{\sqrt{2}} \begin{pmatrix} -\mathrm{i} & -1 \\ \mathrm{i} & -1 \end{pmatrix} \otimes I_{3N}$. $H_2$ can be block-diagonalized

$$\bar{H}_2 = U_2^{\dagger} H_2 U_2 = U_2^{\dagger} \begin{pmatrix} 0 & H_{\mathrm{OBC}} \\ -H_{\mathrm{OBC}} & 0 \end{pmatrix} U_2 = H_3 \oplus (-H_3),$$

meaning that it is separable into two decoupled subsystems, $H_3$ and $-H_3$. The transformation is enabled by $U_2$, also called a shuffle matrix, whose entries $a_{m,n}$ meet the conditions $\begin{cases} a_{m=n} = 1, for \ \mathrm{mod}[n,3] \neq 1 \\ a_{m\neq n} = 1, \mathrm{for\ mod}[\min[m,n],3] = 1 \mathrm{\ and\ } |m-n| = 3N \end{cases}$, $N$ is the number of unit cells in $H_{\mathrm{OBC}}$ ($3N \times 3N$ in dimensions, and $N = 1$ for bulk Hamiltonian).

Because every transformation is bijective (depicted in Fig. 6), properties of $H_{\mathrm{OBC}}$ are preserved throughout. Importantly, the spectrum of $H_3$ is the spectrum of $H_{\mathrm{OBC}}$ rotated by 90° about the zero-energy point, so the zero modes (flat band) in $H_{\mathrm{OBC}}$ are mapped to the zero modes in $H_3$. We can probe the properties of the flat band in $H_3$ without the interference from the imaginary-energy modes in the continuous band. The lattice corresponding to $H_3$ is shown in Fig. 6(a2).

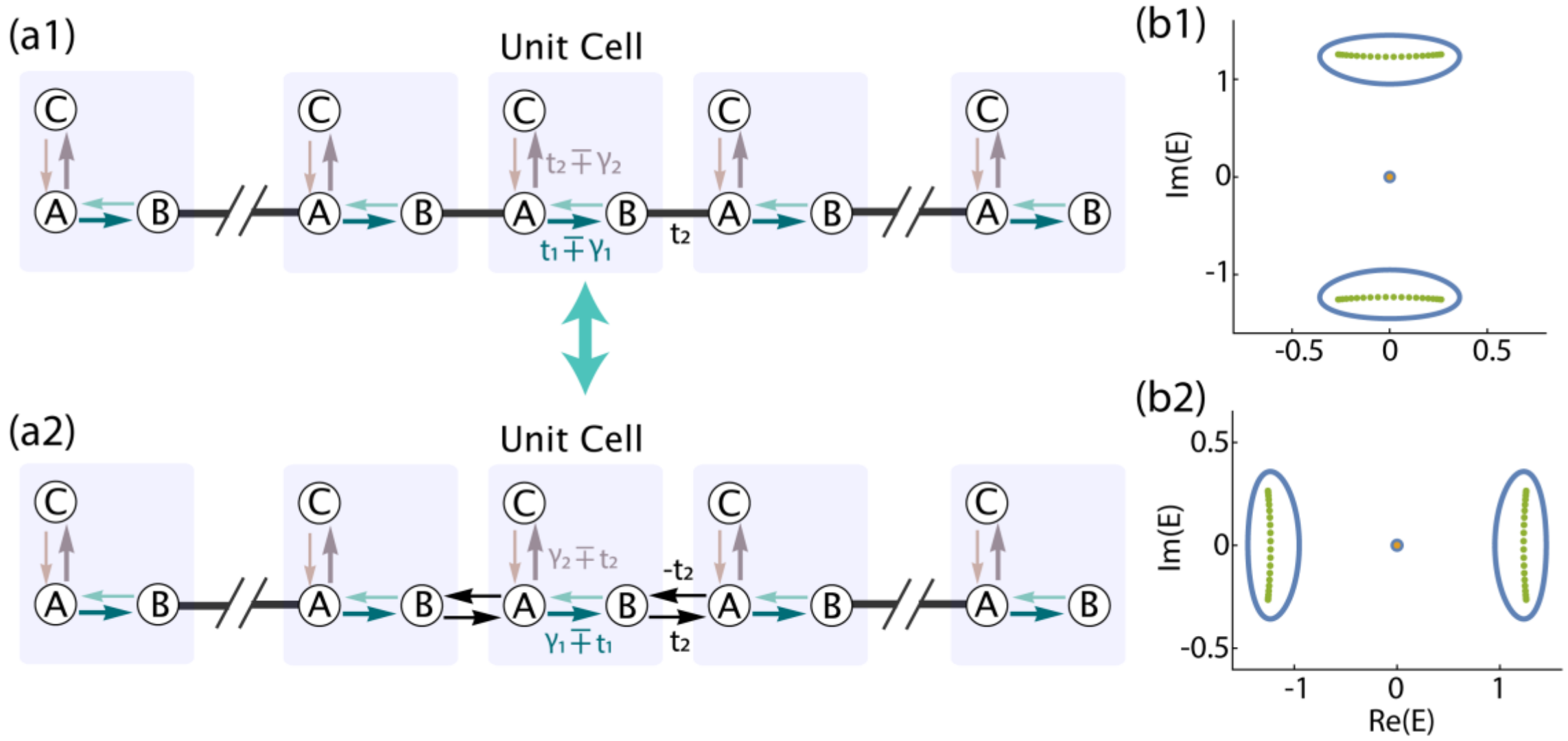

FIG. 6 Transformation of the lattice model. The original lattice (a1) and its spectrum after the reopening of the PBC line gap (b1). The transformed lattice (a2) and its spectra (b2). The spectrum in (b2) is a 90°-rotation of the one in (b1).

*The order of EPs on the OBC flat band*—Special care is needed when determining the order of the EPs on the flat band under OBC. Common methods for determining the order of the EP only detect eigenvalue degeneracy, such as computing the conditions for the vanishing of the characteristic polynomial of the determinant, the resultants of the Hamiltonian. They are no longer reliable because the flat-band eigenmodes are already $N$-fold degenerate even without any EP. In this circumstance, geometric multiplicity is required to correctly identify the EPs.

We first rewrite the OBC Hamiltonian, Eq. (2), under the basis $\boldsymbol{\Psi} = (a_1 \cdots a_N, b_1 \cdots b_N, c_1 \cdots c_N)^{\mathrm{T}}$,

$$H_{\mathrm{OBC}} = \boldsymbol{\Psi}^{\dagger} \begin{pmatrix} 0_{N\times N} & A_{N\times 2N} \\ B_{2N\times N} & 0_{2N\times 2N} \end{pmatrix} \boldsymbol{\Psi} \tag{9}$$

The number of linearly independent zero modes is $N_0 = \dim(\ker H_{\mathrm{OBC}}) = \dim(\ker A) + \dim(\ker B)$. According to the rank-nullity theorem, $\mathrm{col} H_{\mathrm{OBC}} = \mathrm{rank}(H_{\mathrm{OBC}}) + \dim(\ker H_{\mathrm{OBC}})$, where $\mathrm{col}\, H_{\mathrm{OBC}}$ is the number of columns of $H_{\mathrm{OBC}}$. Unless $\gamma_2 = 0$ and $(t_1 + \gamma_1)(t_1 - \gamma_1) = 0$, both $A$ and $B$ are full rank. This means $\dim(\ker B) = \mathrm{col}\, B - \mathrm{rank}\, B = 0$ and $\dim(\ker A) = N_{\mathrm{col}}(A) - \mathrm{rank}\, A = N$. So $\dim(\ker H_{\mathrm{OBC}}) = \dim(\ker A) = N$. In other words, $H_{\mathrm{OBC}}$ always has $N$ linearly independent zero modes, i.e., the geometric multiplicity of the null space is fixed at $N$. Due to the sublattice symmetry of the system, modes on the dispersive bands are always symmetric about the zero energy. So, the modes on the OBC dispersive bands merge to the flat band in a pairwise manner [e.g., see Figs. 4(b1-2)], increasing the algebraic multiplicity at zero energy to $N+2$.

If $t_1+\gamma_1$ and $t_2+\gamma_2$ are simultaneously zero, then $\operatorname{rank} A = N-1$, and $\dim(\ker A) = N_{\mathrm{col}}(A) - \operatorname{rank} A = N+1$. So $H_{\mathrm{OBC}}$ has $N+1$ linearly independent zero modes. In this special case, we can analytically obtain the Jordan canonical form of $H_{\mathrm{OBC}} = Z \oplus D_+ \oplus D_-$, where block $Z$ is $(N+2)\times(N+2)$ in dimension and is associated with the flat band: all its elements are zero, except $Z_{N+1,N+2}=1$. Therefore, the algebraic multiplicity at zero energy is $N+2$ and the EP on the flat band is order-2. This EP2 is formed by the coalescence of two modes from the dispersive bands at zero energy without involving modes from the flat band. An easy way to understand this strange emergence of the EP2 is to observe that $t_2+\gamma_2=0$ means the hopping between sites $C$ and $B$ is purely unidirectional and it hops from $B$ to $C$. On the other hand, blocks $D_{+,-}$ are $(N-1)\times(N-1)$ in dimension and they are associated with the two dispersive bands. However, they are both in a Jordan canonical form with $t_2$ and $-t_2$ on their diagonals, and the sub-diagonal elements are all 1. Therefore, all modes in both dispersive bands form order-$(N-1)$ EPs with energies at $\pm t_2$.

*Experimental setup*—Our experimental realization of a non-Hermitian mechanical lattice comprises three fundamental components: (i) rotational oscillators serving as onsite orbitals, each integrating a brushless DC motor and a rigid arm connected to two fixed anchors by pre-tensioned springs; (ii) reciprocal couplings implemented by two parallel pre-tensioned springs spanning adjacent arms, realizing the nearest-neighbor hoppings $t_1$ and $t_2$; (iii) an active feedback module that generates non-reciprocal coupling coefficients. All oscillators are precisely tuned to the resonant frequency $f_0 = 10.9\ \mathrm{Hz}$ through calibrated mass loading and appropriate spring selection/adjustment. Angular displacements are measured at a sampling rate of $500\ \mathrm{Hz}$ by magnetic rotary encoders (AS5047P) mounted at the motor bases and streamed to a computer via a serial interface.

The non-reciprocal hopping is implemented by a real-time feedback system combining a microcontroller (MCU: Espressif ESP32) and field-oriented control (FOC) drivers (driver chip: STMicroelectronics L6234PD). The active component consists of a brushless motor with a radial magnet rigidly affixed to its base. A magnetic rotary encoder monitors the orientation of the magnet, thereby extracting the angular displacement $\theta$ in real time. This data is to the MCU, which computes the required non-reciprocal torque and drives the adjacent motor. The end-to-end active feedback latency is to $0.35$ ms, which is negligible because the period of $T_0 = 1/f_0 \approx$ $92$ ms.

A chain of 36 oscillators forms 12 unit cells under OBC. Each unit comprises three oscillators and an auxiliary anchor point used to fine-tune the on-site resonant frequencies, thereby ensuring a uniform $f_0$ across all sites within each cell. Consider the non-reciprocal coupling from site A to site B in the $n$-th unit cell: the MCU obtains site-$A$'s instantaneous angular displacement $\theta_{n,A}$ via the encoder, and computes the torque $\tau_{n,B} = \alpha\theta_{n,A}$ (where $\alpha$ is the programmable coupling strength) and drives the motor at site-$B$ through the FOC interface. This implements the non-reciprocal hopping $\gamma_1$. The parameter $\gamma_2$ is implemented analogously. All coupling coefficients are retrieved via the Green's function method.

# Supplemental Material

# Observation of flat-band skin effect

Xulong Wang[1], Dongyi Wang[1], Congwei Lu[1], Ruo-Yang Zhang[2],
Ching Hua Lee[3], Kun Ding[4], Guancong Ma[1,5]

[1]Department of Physics, Hong Kong Baptist University, Kowloon Tong, Hong Kong, China
[2]Department of Physics, Nanjing University, Nanjing, China
[3]Department of Physics, National Singapore University, Singapore 117542, Singapore
[4]Department of Physics, State Key Laboratory of Surface Physics, and Key Laboratory of Micro and Nano Photonic Structures (Ministry of Education), Fudan University, Shanghai 200438, China
[5]Shenzhen Institute for Research and Continuing Education, Hong Kong Baptist University, Shenzhen 518000, China

## 1. The winding of the Gram function and the dispersive bands

As established in the main text (End Matter), the FBSE hinges on the winding of the Gram function $\mathcal{g}(k)$, which is connected to the spectral winding of the dispersive bands with respect to the flat band. Here, we derive this equivalence for an arbitrary 1D $(M+1)$-band non-Hermitian flat-band model $H(k)$. In the following derivation, we consider Bloch Hamiltonian and Bloch wavefunctions, so the variable $k$ is omitted.

Recall that rank $H = M$, so $H$ hosts a flat band at zero energy alongside $M$ dispersive bands. We introduce the adjugate matrix of $H$, denoted as $\mathrm{adj}H$, which satisfies:

$$H \cdot \mathrm{adj}H = \mathrm{adj}H \cdot H = \det H \cdot I. \qquad \text{S1}$$

Because rank $H = M < M + 1$, the determinant vanishes. Consequently, Eq. S1 reduces to

$$H \cdot \mathrm{adj}H = \mathrm{adj}H \cdot H = 0 \qquad \text{S2}$$

This constraint dictates that the column vectors of $\mathrm{adj}H$ must lie entirely in the right null space of $H$ (thus corresponding to the REV of the flat band), while its row vectors reside in the left null space (corresponding to the LEV).

For an $(M+1) \times (M+1)$ matrix has rank $M$, the rank of its adjugate matrix is exactly 1. Therefore, $\mathrm{adj}H$can be decomposed into the outer product of a single column vector and a single row vector:

$$\mathrm{adj}H = \mathrm{const} \cdot |\psi_0^R\rangle\langle\psi_0^L| \qquad \text{S3}$$

where $\langle\psi_0^L|$ and $|\psi_0^R\rangle$ are the unnormalized Bloch LEV and REV of the flat band, and $\mathrm{const}$ is a non-zero scalar factor related to the gauge choice. For our explicit $3 \times 3$ model in the main text, analytical verification yields a trivial constant $\mathrm{const} = -1$. Taking the matrix trace of both sides of Eq. S3, we directly extract the Gram function $\mathcal{g}(k)$:

$$\mathrm{Tr}\ \mathrm{adj}H = \mathrm{const} \cdot \langle\psi_0^L|\psi_0^R\rangle = \mathrm{const} \cdot \mathcal{g}(k) \qquad \text{S4}$$

Simultaneously, from the perspective of characteristic polynomials, the trace of an $(M+1) \times (M+1)$ adjugate matrix, $\mathrm{adj}H$, equals the sum of all its principal minors of order $M$. By the Vieta's formulas, this sum is identical to the sum of all possible $M$-fold products of its $M+1$ eigenvalues:

$$\mathrm{Tr}\ \mathrm{adj}H = \sum_{i=1}^{M}\left(\prod_{j\neq i} E_j(k)\right) \tag{S5}$$

Due to the zero-energy flat band, any product combination containing these zero eigenvalues vanishes. The sum therefore collapses to the product of the $M$ non-zero dispersive band energies

$$\mathrm{Tr}\ \mathrm{adj}H = \prod_{j=1}^{M} E_j(k) \tag{S6}$$

By equating Eqs. S4 and S6, we arrive at the identity connecting the Gram function $g(k)$ to the dispersive energy spectrum

$$g(k) = \mathrm{const}\prod_{j=1}^{M} E_j(k). \tag{S7}$$

According to the definition of the non-Hermitian winding number, $\mathcal{W} = \frac{1}{2\pi i}\int_{\mathrm{BZ}} dk\,\ln(E(k) - E_{\mathrm{ref}})$, the winding number of a product of functions mathematically factorizes into the sum of their individual winding numbers. The winding number of the Gram function $g(k)$ identically equals the sum of the winding numbers of all dispersive bands

$$\mathcal{W}_{g} = \sum_{j=1}^{M} \mathcal{W}_{E_j} \tag{S8}$$

Specifically, regarding the 3-level flat band model ($M = 2$) in the main text, $g(k) = -E_1(k)E_2(k)$, where $E_1(k)$ and $E_2(k)$ are the energy of two dispersive bands.

In conclusion, the winding of the dispersive bands around the flat band is equivalent to the winding of the Gram function $g(k)$, which then determines the "NHSE" in the Gram matrix $\mathcal{G}$ that gives rise to the FBSE. In Fig. S1, we can see that the winding of $g(k)$ exactly reproduces the phase diagram of the spatially averaged response $\chi$ computed numerically in the main text.

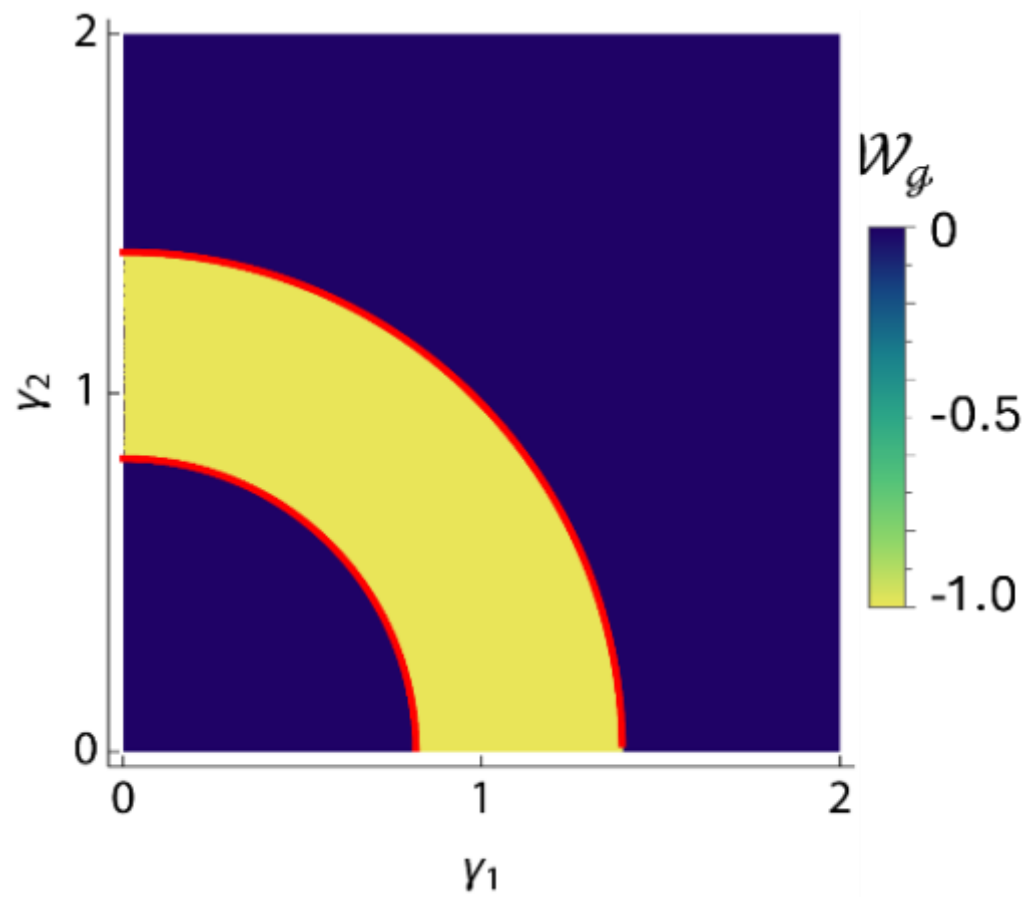

FIG. S1. The winding number $\mathcal{W}_g$ computed via Gram function $g(k)$ with $t_1 = -1.06$, $t_2 = -0.3$.

## 2. Green's function indicator for the flat-band skin effect

Non-Hermitian skin effects (NHSE) are typically diagnosed by PBC spectral winding or by the direct observation of the edge-localized skin-mode wavefunctions. But these indicators are unfit for FBSE. As discussed in our paper, flat bands clearly have no winding. Also, owing to the high degeneracy of the OBC flat-band modes, any wavefunction profile can be constructed through their linear combinations. This is true even if the FBSE is absent, i.e., one can construct edge-localized mode through a specific linear combination of CLS. Therefore, any direct observation of a single skin-mode-like wavefunction is inconclusive for flat bands. Similarly, other indicators relying on real-space wavefunctions, such as the inverse participation ratio (IPR), face the same basis-dependency issue. The Lyapunov exponent is also unsuitable because it cannot readily distinguish between bulk-distributed CLS and boundary-localized skin modes.

This is why we introduced a new indicator for diagnosing the FBSE. The object $\chi$ derived from the Green's function perfectly serves this purpose because: (1) it directly corresponds to the macroscopic steady-state responses observable in experiments; and (2) it disregards the basis chosen for the OBC flat band by involving all modes.

### 2.1 Scaling of Green's function

It has been established that under NHSE, the maximum entry of the Green's function, $\hat{G}(E_\eta) = (E_\eta - H_{\mathrm{OBC}})^{-1}$, scales exponentially with the lattice size [1]. To demonstrate this behavior for the FBSE, we analyze the 1D non-Hermitian flat-band tight-binding model governed by Eq. (2) in the main text. The system's responses are shown in Fig. S2 as functions of lattice size $N$. Therein, it is seen that the maximum entries of the Green's functions indeed scale with $N$ when FBSE is present [Fig. S2(b)], but are insensitive to $N$ when FBSE is absent [Fig. S2(a, c)], where the flat band is outside the point gaps of the dispersive bands under periodical boundary condition (PBC).

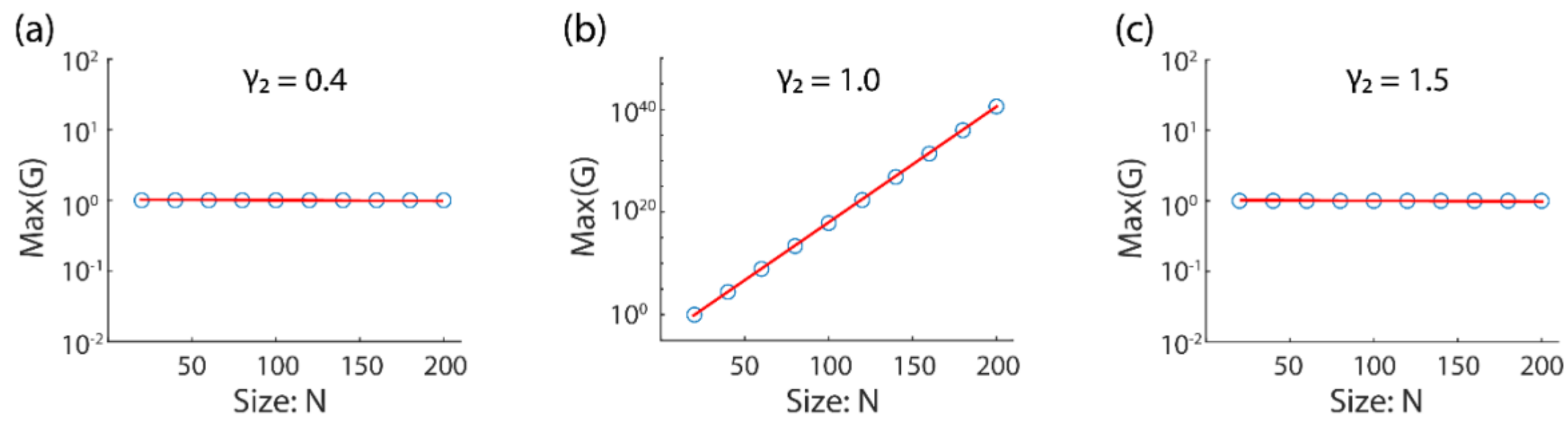

FIG. S2. Scaling of the maximum entry of the flat band Green's function $G(E_\eta)$. No scaling is seen in (a, c), at which the PBC spectrum of the flat band is outside of the point gap of the dispersive bands. Exponential scaling is seen in (b), at which the flat band is inside the point gap. Other parameters are $t_1 = -1.06$, $t_2 = -0.3$ and $\gamma_1 = 0.5$.

**2.2 Robustness of the Green's function indicator**

To demonstrate the robustness of this indicator, we evaluate the sensitivity of the spatially averaged response ($\chi$) against various excitation conditions (e.g., a Gaussian wave packet and a square wave distributed over five sites) and different spatial positions of the point source (e.g., exciting site-C in the 5th, 7th, and 20th unit cells) in a lattice comprising 100 unit cells ($t_1 = -1.06$, $t_2 = -0.3$), as shown in Figs. S3 and S4.

In the Green's function, $G(E_\eta) = \sum_{n=1}^{N} \frac{|\psi_{0,n}^R\rangle\langle\bar{\psi}_{0,n}^L|}{E_\eta - \omega_n}$, taking the limit $E_\eta \to 0$ is essential to make sure that the response is overwhelmingly dominated by the flat-band modes. We further investigate the sensitivity of $\chi$ by approaching $E_\eta \to 0$ from both the real axis ($E_\eta \in [10^{-3}, 10^{-5}, 10^{-10}]$) and the imaginary axis ($E_\eta \in [10^{-3}\mathrm{i}, 10^{-5}\mathrm{i}, 10^{-10}\mathrm{i}]$), as depicted in Fig. S5. These results confirm that the Green's function serves as a robust indicator of the FBSE.

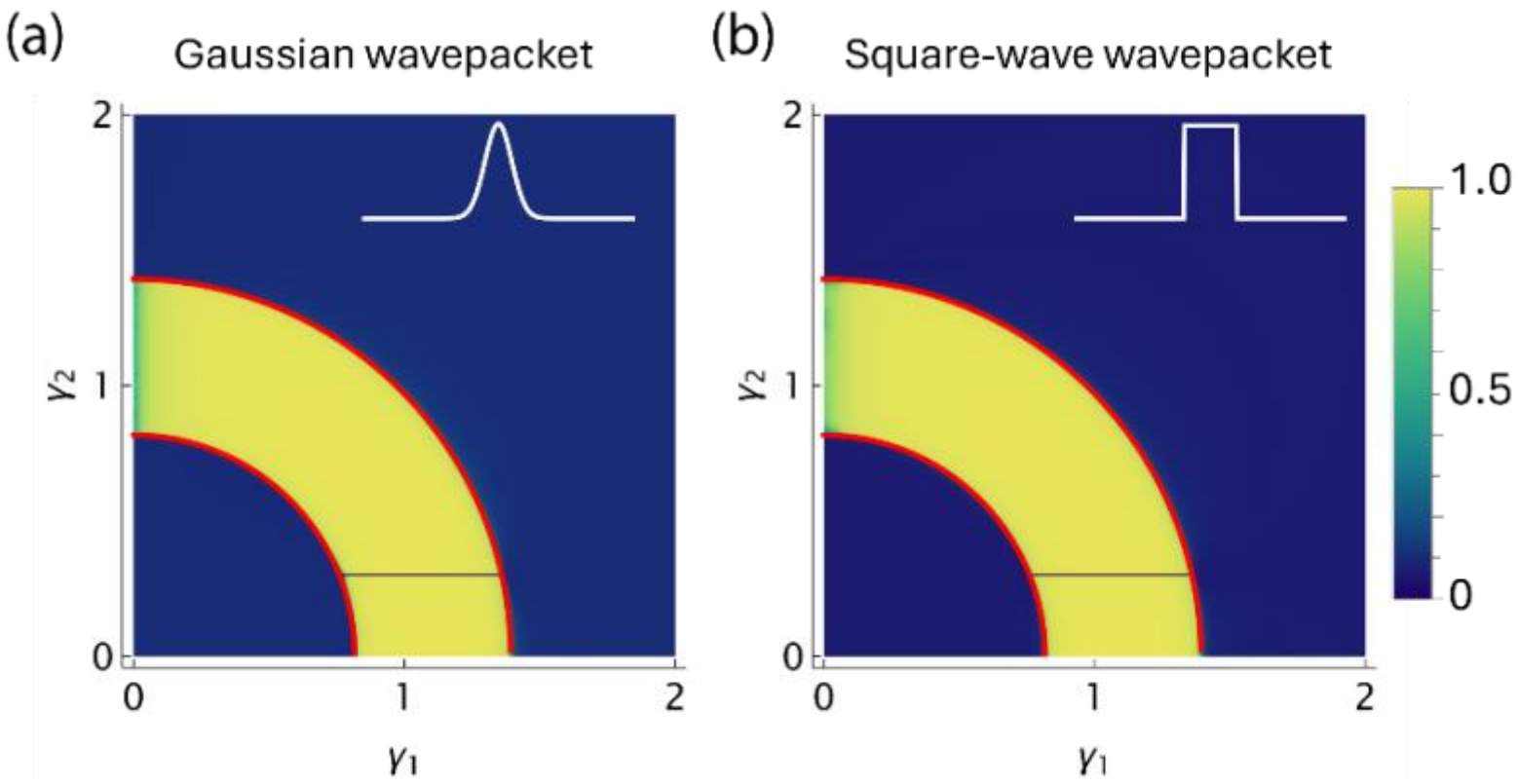

FIG. S3. The averaged responses ($\chi$) driven by different source profiles (Gaussian wave packet and square wave distributed over five sites). Two profiles reproduce the same FBSE phase diagram.

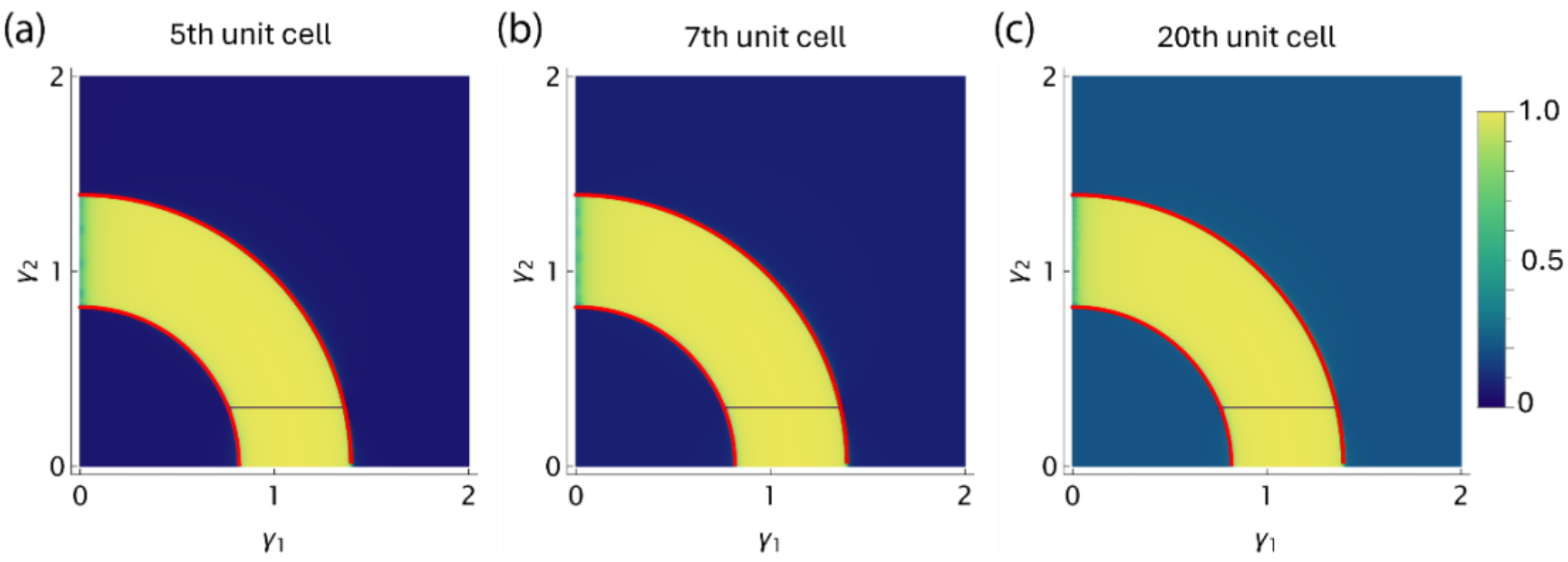

FIG. S4. The averaged responses ($\chi$) under varying excitation locations (exciting site C in the 5th, 7th, and 20th unit cells). As the excitation location moves toward the center of the lattice, the averaged response ($\chi$) consistently approaches to 1 (for right-localized FBSE) in the FBSE region.

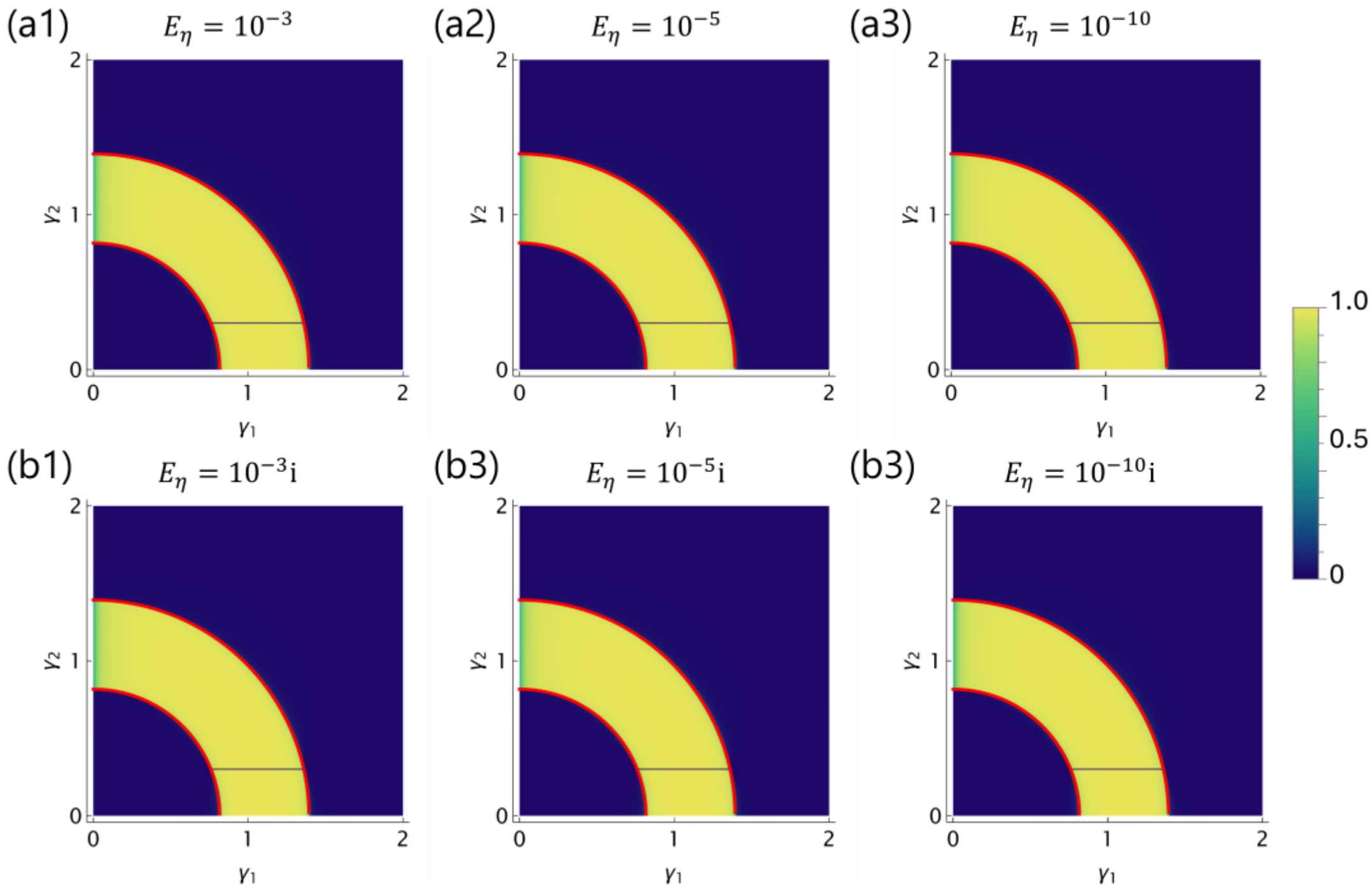

FIG. S5. The averaged responses $\chi$ evaluated under different limits ($E_\eta \to 0$ along imaginary and real axes). The numerical results show the same FBSE region with nearly invariant $\chi$.

## 3. Effects of symmetry-breaking perturbations on the flat band

The manifestation of the FBSE relies on the rank-deficient structure of the Hamiltonian. This structure is broken when a small perturbation, such as a constant on-site potential, is introduced. As such, the flat band is no longer flat and it acquires a dispersion. Once dispersion is introduced, the conventional point-gap topology becomes well-defined for this band. As a result, the FBSE transitions into the conventional NHSE. We numerically verified this by applying the constant perturbation $\delta V = 0.01$ on sites C. The results are shown in Fig. S6.

In Figs. S6 (a-c), the original flat band lies outside the point gaps of the dispersive bands, so FBSE does not appear. Under the perturbation, the original flat band forms a small point gap that encloses the corresponding OBC modes [Figs. S6(a, b)]. NHSE is seen for these modes [Fig. S6(c)]. The similar is observed for the case when both non-Hermitian parameters are large and imaginary line gaps separate the dispersive and the flat bands, as shown in Figs. S6(g-i).

In Figs. S6(d-f), the original flat band is inside the point gap formed by the two dispersive bands. The perturbed flat band is seen to merge with a dispersive band in the formation of a combined spectral loop that encloses the OBC states [Figs. S6(d, e)]. NHSE appears for the OBC lattice [Fig. S6(f)].

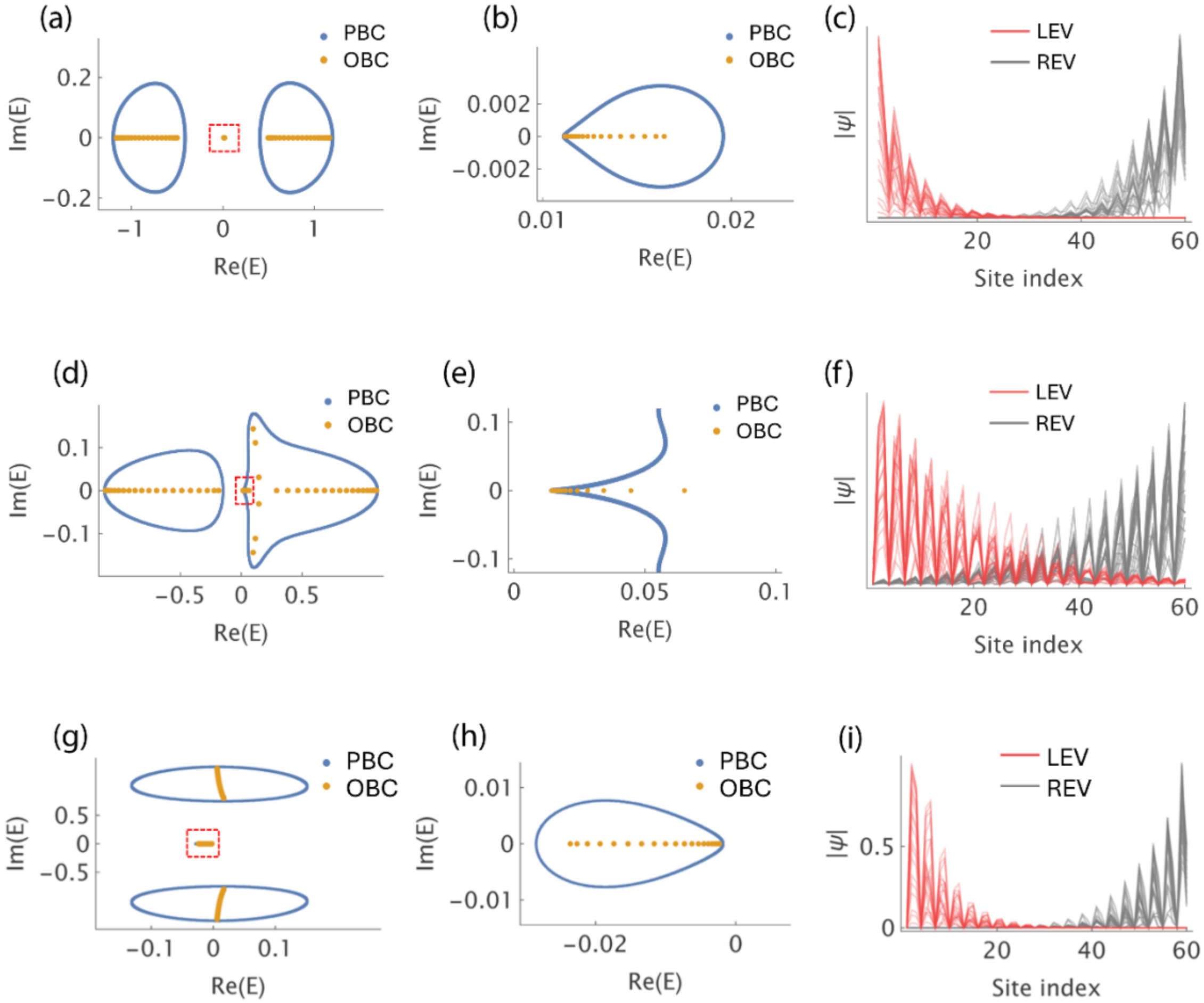

FIG. S6. (a, d g) The complex energy spectra of the system under PBC (blue lines) and OBC (orange dots) with a weak on-site perturbation $\delta V = 0.01$. The perturbed flat band is located outside (a, g) and inside (d) the point gaps formed by the dispersive bands. (b, e, h) Enlarged view of the red dashed box in (a, d, g). By acquiring a weak dispersion due to the perturbation, the original flat band forms a distinct spectral loop in the complex energy plane, completely enclosing the corresponding OBC states. (c, f, i) Spatial distributions of LEVs (red lines) and REVs (grey lines) corresponding to the modes in (a, d, g). LEVs and REVs are exponentially localized at the left and right boundaries, respectively, explicitly demonstrating the recovery of the conventional NHSE. Parameters are set to $\gamma_1 = 0.5, \gamma_2 = 0.5$ for (a-c), $\gamma_1 = 0.2, \gamma_2 = 0.8$ for (d-f) and $\gamma_1 = 0.5, \gamma_2 = 1.5$ for (g-i).

## 4. Third-order exceptional points and their observable consequences

In our non-Hermitian model, flat band and dispersive bands coalesce at third-order exceptional points (EP3s). Consequently, the Green's function, $G(\omega) = (\omega - H)^{-1}$, has higher-order poles proportional to $\omega^{-3}$. In comparison, the flat band exhibit first-order poles ($\omega^{-1}$) away from the EP3. To show this, we calculate the steady-state response of the OBC lattice with 6 unit cells under two parameter sets. We applied a local harmonic excitation at site B of the first unit cell and measured the response amplitude at site B ($|R_2|$), sweeping the excitation frequency $\omega$ approaching the flat band energy. As shown in Fig. S7a (plotted on a log-log scale), the response amplitude exactly follows $\omega^{-3}$ at EP3s (with a slope of $-3$ in the log-log plot). In contrast, when the system parameters are tuned away from the EP3, the response strictly follows the conventional $\omega^{-1}$ scaling (with a slope of $-1$ in the log-log plot).

Furthermore, the response at EP3 has a different line shape [2]. As illustrated in Fig. S7b, when a uniform background loss (implemented by adding $0.01\mathrm{i}$ to the Hamiltonian) is introduced to the lattice, the response spectrum around the zero energy exhibits a significantly sharper resonance peak at the EP3 compared to the typical flat-band response without the EP. This enhanced frequency sensitivity, and the distinct scaling behavior serve as direct observable consequences of EP3s in our model.

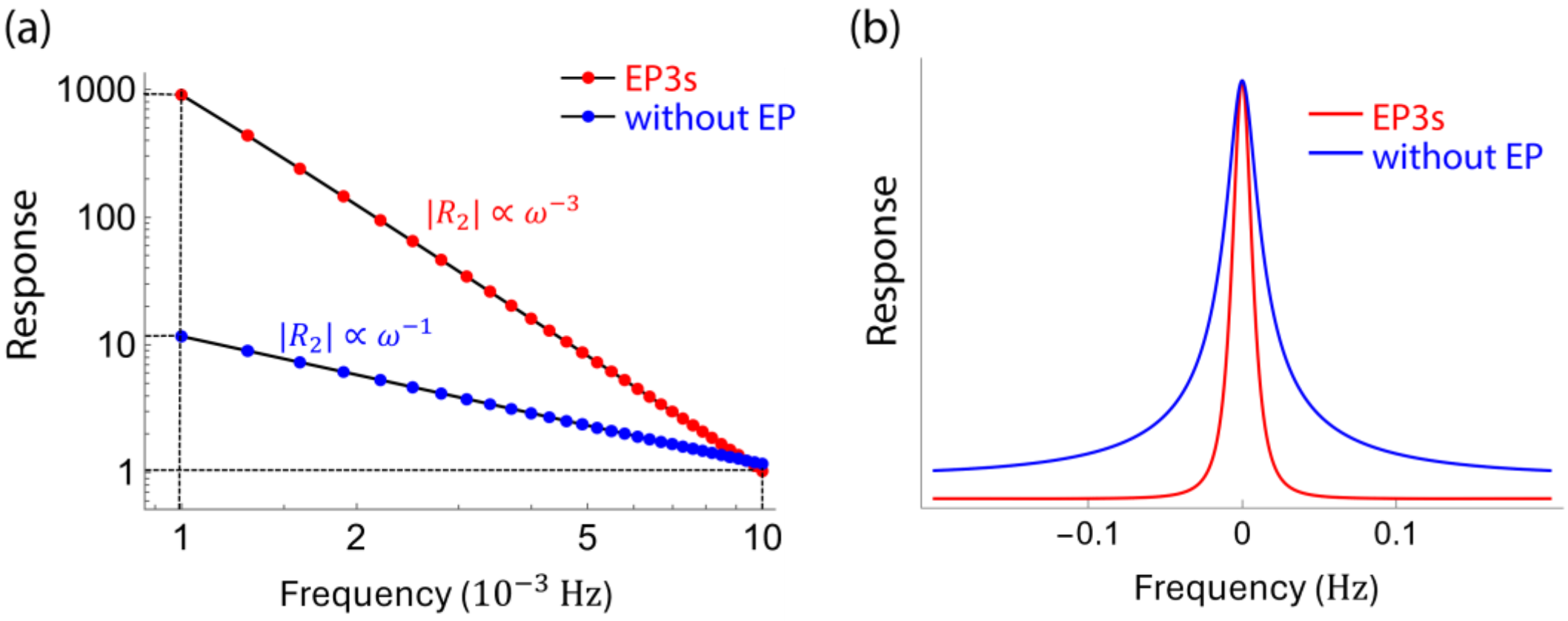

FIG. S7. Observable consequences of EP3s under open boundary conditions. (a) Frequency scaling of the steady-state response amplitude $|R_2|$ plotted on a log-log scale. The red and blue dots represent the numerically calculated responses at the EP3 ($\gamma_1 = 0.5$ and $\gamma_2 = 0.736286$) and away from the EP (without EP, $\gamma_1 = 0.5$ and $\gamma_2 = 0.5$), respectively. The solid lines indicate the theoretical scaling laws of $\omega^{-3}$ and $\omega^{-1}$. To clearly visualize and compare the distinct slopes within the same frame, the plotted response values have been rescaled by constant factors (divided by $2 \times 10^4$ for the EP3 case and divided by 5 for the non-EP case). (b) The steady-state response spectrum around zero energy in the presence of a uniform background loss ($0.01 * 1$i). Due to the higher-order pole, the system exhibits a significantly sharper resonance peak at the EP3 (red curve) compared to the conventional response without the EP (blue curve).

## 5. Flat-band skin effects in other flat band models

The FBSE is a generic phenomenon in non-Hermitian flat-band models. Here, we show its emergence in other paradigmatic flat-band models, including a non-Hermitian AB cage lattice (Fig. S8), a ladder model (Fig. S9), and a Lieb lattice (Fig. S10). The results consistently show that the existence of FBSE is dictated by the point-gap topology of the dispersive bands. The key condition for the emergence of the FBSE is that the energy of the flat band lies inside the point gap of the dispersive-band spectrum under PBC.

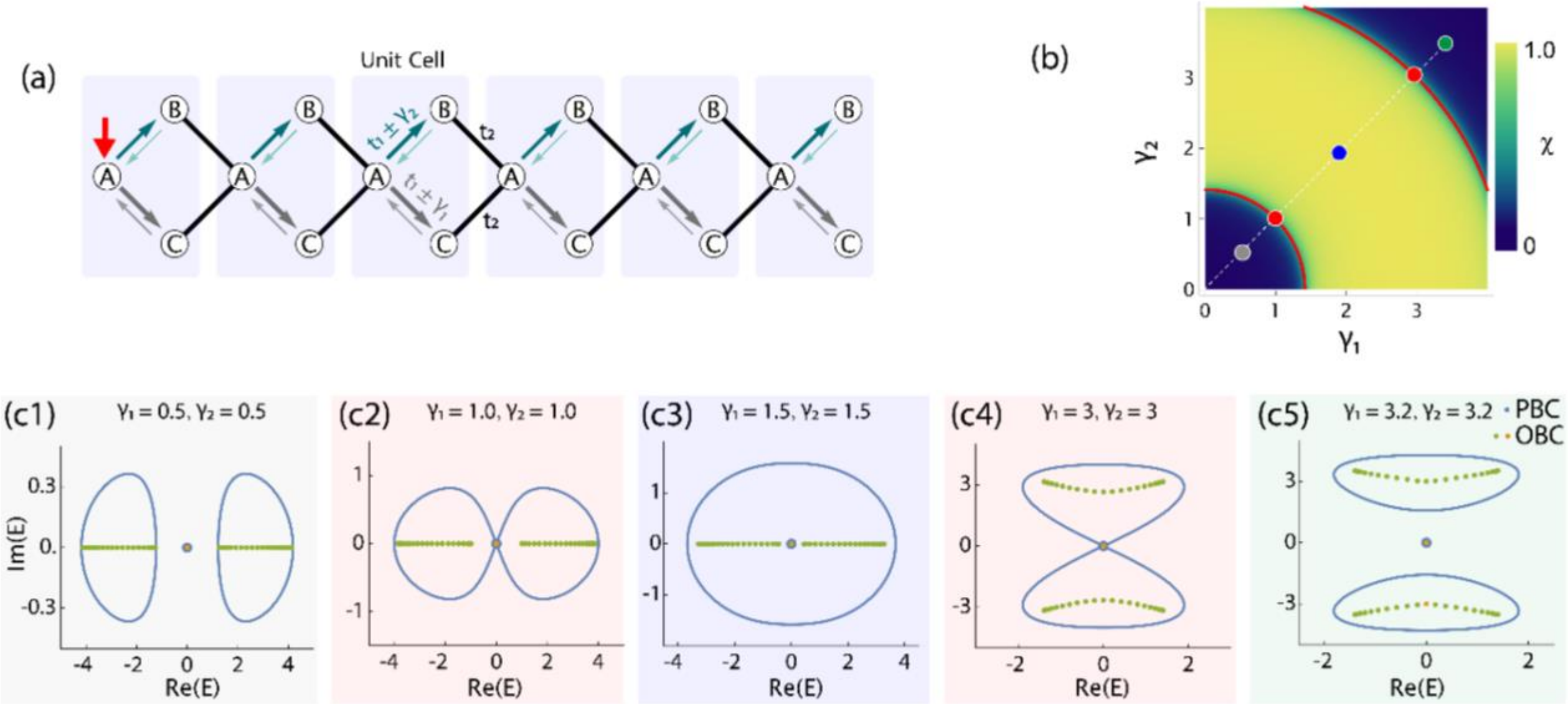

FIG. S8. (a) A 1D non-Hermitian AB cage model. (b) $\chi$ of the OBC flat band computed through the Green's function. The excitation position is at the leftmost A-site as marked by the red arrow in (a). The red curves mark the parametric positions where the PBC line gap of the dispersive bands closes and reopens. (c1-c5) The spectra for both PBC (blue) and OBC (green and orange)

at five representative parameter sets, marked by the colored circles in (b). We set parameters to $t_1 = 2$ and $t_2 = 1$.

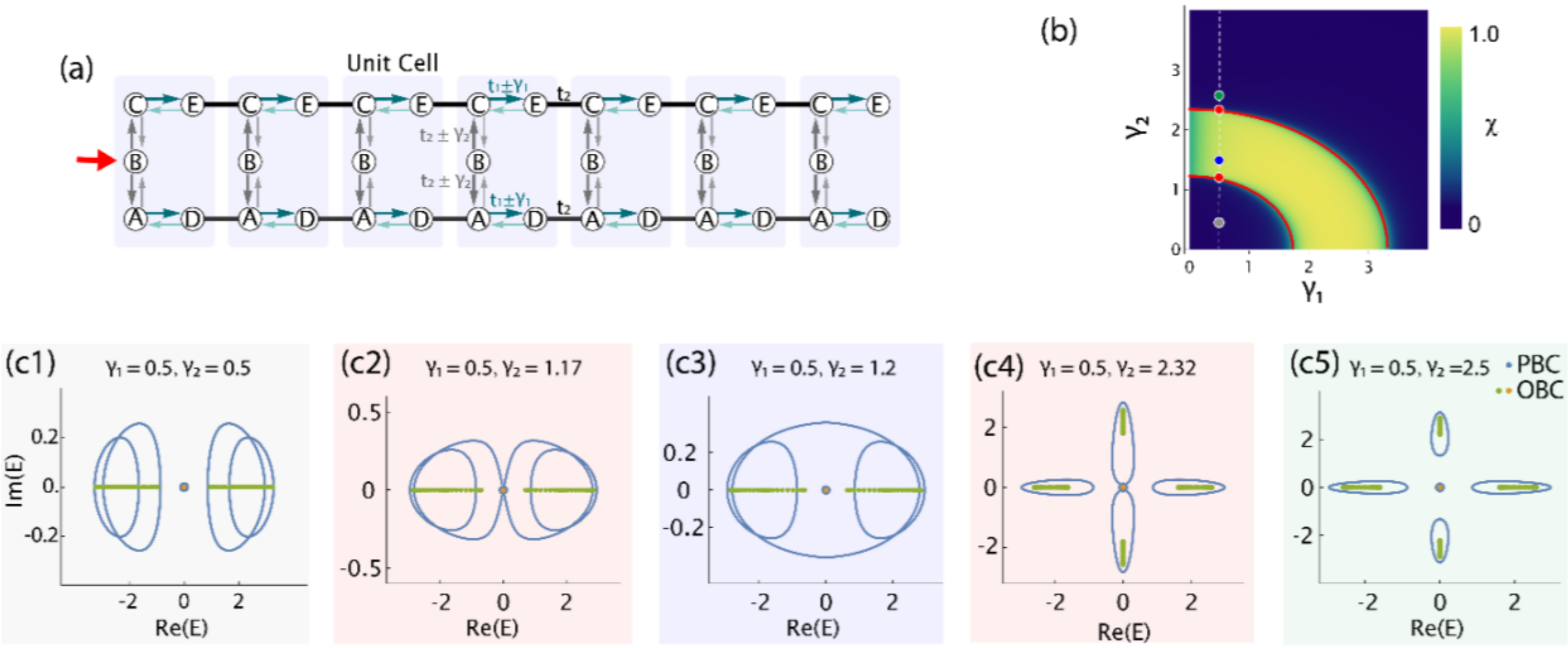

FIG. S9. (a) A 1D non-Hermitian ladder model. (b) $\chi$ of the OBC flat band computed through the Green's function. The excitation position is at the leftmost B-site as marked by the red arrow in (a). The red curves mark the parametric positions where the PBC line gap of the dispersive bands closes and reopens. (c1-c5) Spectra for both PBC (blue) and OBC (green and orange) at five representative parameter sets, marked by the colored circles in (b). We set parameters to $t_1 = 2$ and $t_2 = 1$.

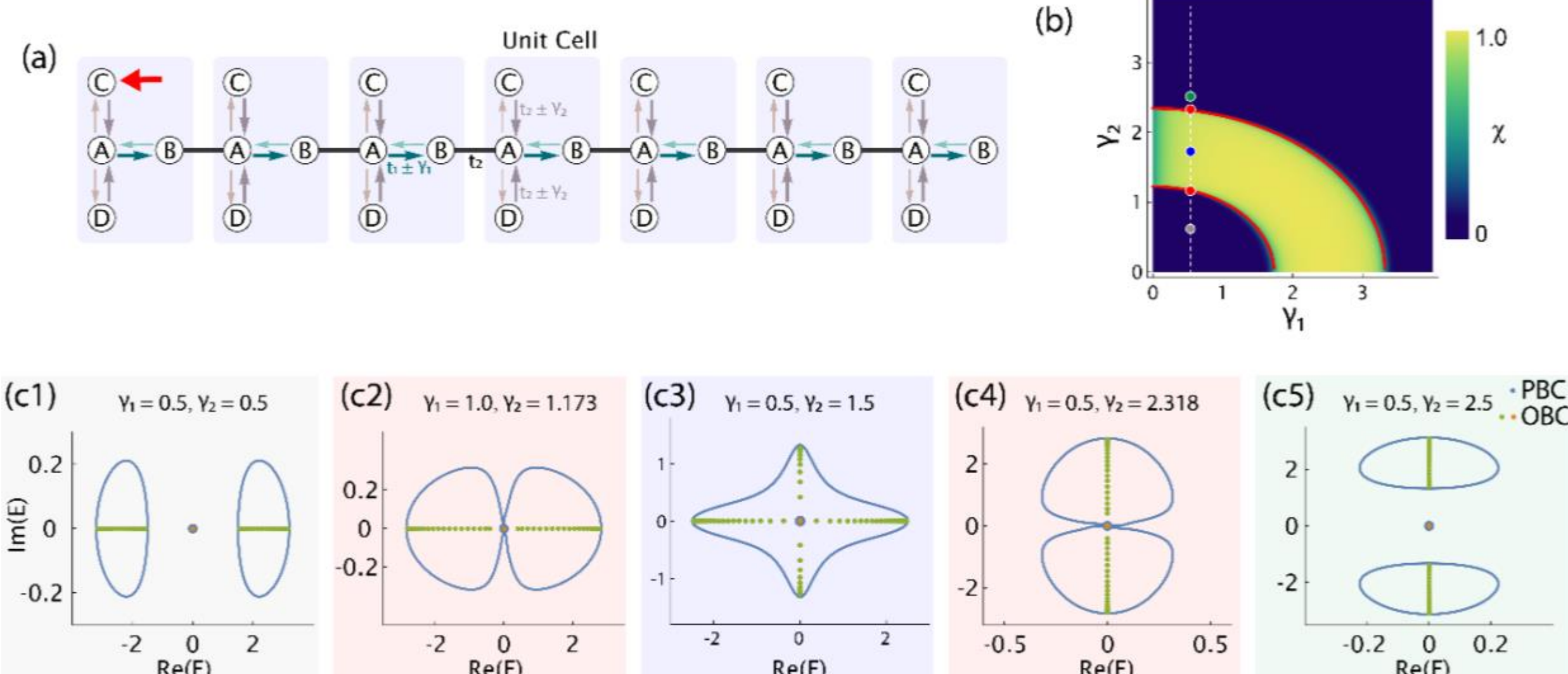

FIG. S10. (a) A 1D non-Hermitian Lieb model. (b) $\chi$ of the OBC flat band computed through the Green's function. The excitation position is at the leftmost C-site as marked by the red arrow in (a). The red curves mark the parametric positions where the PBC line gap of the dispersive bands closes and reopens. (c1-c5) Spectra for both PBC (blue) and OBC (green and orange) at five representative parameter sets, marked by the colored circles in (b). We set parameters to $t_1 = 2$ and $t_2 = 1$.